\documentclass{aa} 
\usepackage{txfonts}  
\usepackage{graphicx}  
\usepackage{longtable,lscape}
\usepackage{natbib}  
\usepackage[normalem]{ulem}
\bibpunct{(}{)}{;}{a}{}{,} 
\newcommand{\nc}{\newcommand}  
\newcommand{\micron}{\mbox{$\mu$m}}  
\newcommand{\msun}{M$_{\rm \odot}$}  
\newcommand{\lsun}{L$_{\rm \odot}$}

\newcommand{\HII}{\ion{H}{ii}}  
\newcommand{\av}{$A_{\rm V}$}  
\newcommand\arcdeg{\mbox{$^\circ$}}%
\newcommand{\tdust}{\mbox{$T_{\rm d}$}}

\newcommand{\tex}{\mbox{$T_{\rm ex}$}}
\nc{\twCO}{$^{12}$CO}  
\nc{\thCO}{$^{13}$CO}  
\nc{\CeiO}{C$^{18}$O}  
\nc{\COseventeen}{C$^{17}$O}  
\nc{\cmcub}{\mbox{cm$^{-3}$}}  
\nc{\cmsq}{\mbox{cm$^{-2}$}}  
\nc{\Kkms}{\mbox{K~km~s$^{-1}$}}  
\nc{\kms}{\mbox{km~s$^{-1}$}}  
\nc{\ext}{\mbox{T$_{\rm ex}$}}  
\nc{\nhtwo}{\mbox{N(H$_2$)}}
\nc{\nhtwod}{\mbox{NH$_2$D}}
\nc{\vlsr}{\mbox{$V_{\rm LSR}$}}
\nc{\Cone}{\mbox{[\ion{C}{i}]$^3$P$_1$--$^3$P$_0$}}  
\nc{\Ctwo}{\mbox{[\ion{C}{i}]$^3$P$_2$--$^3$P$_1$}}  
\nc{\cbyco}{\mbox{C/CO}}
\def\Msun{\,{\rm M$_{\odot}$}}
\def\Lsun{\,{\rm L$_{\odot}$}}
\def\1figthree#1#2#3#4{\begin{figure}[#4]
\hspace*{-3em}
\centering \leavevmode
\epsfxsize=.33\textwidth \epsfbox{#1} \hfil
\epsfxsize=.33\textwidth \epsfbox{#2}\hfil
\epsfxsize=.33\textwidth \epsfbox{#3}
\vspace*{-4pt}
\end{figure}
}

\begin{document}  
  
\title{Young Stars and Protostellar Cores near NGC\,2023} 
    
\author{B. Mookerjea \inst{1}
\and G. Sandell \inst{2} 
\and T. H. Jarrett \inst{3}
\and J. P. McMullin \inst{4,5}
}
  
\institute{ Department of  Astronomy \& Astrophysics, Tata Institute 
of Fundamental Research, Homi Bhabha Road, Mumbai 400005, India
\and SOFIA-USRA, NASA Ames Research Center, MS 211-3,
Moffett Field, CA 94035, USA
\and Spitzer Science Center, California Institute of Technology,
Pasadena, CA, 91125, USA
\and Joint ALMA Observatory, Av Apoquindo 3650 Piso 18, Las Condes, 
Santiago, Chile
\and The National Radio Astronomy Observatory, 520 Edgemont Road
Charlottesville, VA 22903-2475, USA
}

\offprints{B. Mookerjea, \email bhaswati@tifr.res.in}  
  
\date{Received /Accepted }  
  
\titlerunning{YSOs near NGC\,2023 }
\authorrunning{Mookerjea et al.}

\abstract  
{We investigate the young (proto)stellar population in NGC~2023 and
the L~1630 molecular cloud bordering the {{\sc
H~ii}} region IC\,434,  using Spitzer
IRAC and MIPS archive data, JCMT SCUBA imaging and spectroscopy as
well as targeted BIMA observations of one of the Class 0 protostars,
NGC\,2023 MM\,1.  } {We study the distribution of gas, dust and young
stars in this region to see where stars are forming, whether the
expansion of the {{\sc H~ii}} region has triggered star formation, and
whether dense cold cores have already formed stars. } {We have
performed photometry of all IRAC and MIPS images, and used color-color
diagrams to identify and classify all  young stars seen within a
22\arcmin\ $\times$ 26\arcmin\ field along the boundary between
IC\,434 and L\,1630. For some stars, which have sufficient optical,
IR, and/or sub-millimeter data we have also used the online SED
fitting tool for a large 2D archive of axisymmetric radiative transfer
models to perform more detailed modeling of the observed SEDs.  We
identify 5 sub-millimeter cores in our 850 and 450 $\mu$m SCUBA
images, two of which have embedded class 0 or I protostars.
Observations with BIMA are used to refine the position and
characteristics of the Class 0 source NGC\,2023\,MM\,1.  These
observations show that it is embedded in a very cold cloud core, which
is strongly enhanced in NH$_2$D.} {We find that HD\,37903 is the most
massive member of a  cluster with 20 -- 30 PMS stars. We  also find
smaller groups of PMS stars formed from the Horsehead nebula and
another elephant trunk structure to the north of the Horsehead. Star
formation is also  occurring in the dark lane seen in IRAC images and
in the sub-millimeter continuum. We refine the spectral classification
of HD\,37903 to  B2 Ve. We find that the star has a clear IR
excess, and therefore it is a young Herbig Be star.} {Our study shows
that the expansion of the IC\,434 {{\sc H~ii}} region has triggered
star formation in some of the dense elephant trunk structures and
compressed gas  inside the L\,1630 molecular cloud.  This pre-shock
region is seen as a sub-millimeter ridge in which stars have already
formed. The cluster associated with NGC\,2023 is very young, and has a
large fraction of Class I sources.}

\maketitle  
  
\section{Introduction}

The southern part of the Orion B giant molecular cloud complex,
L\,1630, borders the large \HII\ region IC\,434, which is expanding
into the molecular cloud and possibly triggering star formation. The
interface between the molecular cloud and the \HII\ region is seen as
a bright north-south ridge of glowing gas with the Horsehead nebula
(B33) and several smaller ``elephant trunks" seen in silhouette
against the bright nebulosity. B33 points directly towards the binary
system, $\sigma$~Ori, which is ionizing the \HII\ region IC\,434
\citep{abergel2002,Pound03}. Since the Horsehead is seen in absorption
against the nebula, it must be at the same distance or closer to us
than $\sigma$ Ori, which has a distance of  330--385 pc
\citep{Brown94,Caballero08}. Although  350~pc is a more probable
distance, in this paper we adopt the distance 400~pc, which allows
easy comparison to other studies of this region.  The derived
luminosities may therefore be overestimated by 20 to 30\%, and
physical sizes by 10--20\%.

The first systematic search for dense cores and embedded  star
clusters in the L\,1630 cloud complex was carried out by
\citet{Lada91,Lada91a}, who surveyed 3.6 square degrees in CS $J = 2
\to 1$ (98~GHz) and in K-band (2.2 $\mu$m)  down to 14~mag. They
identified 4 embedded clusters,  the richest being associated with
NGC\,2024 with over 300 embedded stars, and the poorest associated
with NGC 2023, a reflection nebula just north of the Horsehead with
only 21 embedded sources. A more detailed study by \citet{depoy1990}
in J, H, and K with a limiting K magnitude of 15 reduced the number of
young stars in the NGC\,2023 cluster to 16.  \citet{launhardt1996}
surveyed the CS J = $2 \to 1$ cores for dust emission at 1.3~mm, and
found most of them to be associated with compact dust emission.

In the vicinity of NGC\,2023 \citet{launhardt1996} found faint
extended emission associated with the southern part of the reflection
nebula; to the west of the nebula they found a relatively bright
double  source LBS\,34\,SM\,1 and SM\,2 embedded in fainter extended
emission. \citet{sandell1999} serendipitously re-discovered the latter
two sources while carrying out a CO $J = 2 \to 1$ survey of the
Horsehead and L\,1630 and found  an extreme high velocity outflow
associated with and driven by LBS\,34\,SM\,1, which they called
NGC\,2023\,MM~1.  \citet{sandell1999} carried out extensive
observations of NGC\,2023\,MM~1 and  MM~2 and found that MM~1 had all
the characteristics of a Class 0 protostar with a luminosity of $\leq$
10 \Lsun.  There are other signposts of star formation in the vicinity
of NGC~2023 such as Herbig-Haro objects \citep{malin1987} and
free-free emission sources \citep{reipurth2004} without any obvious
optical or near-IR counterparts. These observations suggest that there
may be a more embedded population of protostars hidden in the dense
molecular cloud cores surrounding NGC\,2023.

\citet{bowler2009} have recently derived an infrared census of star
formation in a 7\arcmin$\times$7\arcmin\ region around the Horsehead
nebula. They used deep near-infrared (IRSF/SIRIUS $J~H~K_{\rm
s}$) and mid-IR ({\em Spitzer IRAC}) observations. Their findings
support triggered star formation due to radiation driven implosion in
the Horsehead, although they see no evidence for sequential star
formation in the immediate vicinity of the cloud/\HII\ region
interface.  They identify three bona fide and five candidate young
stellar objects (YSOs) in the Horsehead area.

In this paper we report on millimeter  aperture synthesis observations
of NGC\,2023\,MM~1 and MM~2 with BIMA\footnote{The BIMA array was
operated by the Universities of California (Berkeley), Illinois, and
Maryland with support from the National Science Foundation.} to better
characterize the protostars and to see if we could resolve its
accretion disk. We also carried out more extensive continuum mapping
of the NGC\,2023 field with SCUBA on JCMT\footnote{James Clerk Maxwell
Telescope (JCMT)  is operated by the Joint Astronomy Centre on behalf
of the Science and Technology Facilities Council of the United
Kingdom, the Netherlands Organisation for Scientific Research, and the
National Research Council of Canada.} to look for young deeply
embedded stars, which can be too heavily obscured to be detected in
the near infrared (e.g.  NGC\,2023\,MM\,1). We have also retrieved all
IRAC and MIPS images from the archives of Spitzer Space Observatory;
these data go much deeper and to longer wavelengths than those
available from any ground based observatory. The \citet{bowler2009}
study of the Horsehead area uses the same IRAC data as we analyze in
this study, but we examine a much larger field. Since we started this
project, two other SCUBA studies of the Orion B South region have been
published \citep{johnstone2006, nutter2007}. Both these studies are in
part based on the same SCUBA observations. Although we do not go as
deep as \citet{johnstone2006}, we have performed a very careful data
(re-) reduction, to achieve better astrometry, sharper images and well
calibrated data. This has enabled us to identify infrared, radio (and
X-ray) counterparts to the sources detected in our SCUBA observations.

\section{Observation}

\subsection{BIMA observations}
\label{BIMA_obs}
The observations of NGC\,2023\,MM\,1/MM\,2 were made on Dec 13, 1999
and April 13, 2000 using one frequency setting in the C-array
configuration of the  Berkeley-Illinois-Maryland Association (BIMA)
interferometric array. The weather conditions were marginal
in December and somewhat better in April, but both runs had poor phase
stability. The correlator was split into four 25 MHz bands giving a
velocity resolution of  $\sim$0.37 km~s$^{-1}$.  Since the single dish
observations by \citet{sandell1999} indicated that NGC\,2023\,MM\,1
has rather strong emission in deuterated lines, we decided to look for
NH$_2$D 1$_{11} \to  1_{01}$ (at 85.9262703~GHz), which is typically
quite abundant in cold cloud cores. We also included SO  2$_2\to1_1$,
which can probe both the outflow and the cold molecular gas in the core.
We kept the upper sideband free of any molecular transitions that
could be expected to be excited in a cold low mass cloud core. 3C~273
was observed for 10 minutes for bandpass calibration and the quasar
0609$-$157 for phase calibration at intervals of 30 minutes. The
observations of 0609$-$157 were done using a bandwidth of 800 MHz for
2.5 minutes. Since there were no planets available during the
observations we used our phase and bandpass calibrators for flux
calibration. At the time of the observations 3C~273 had a flux density
of 8.8 Jy, while the flux density of 0609$-$157 was 5.0 Jy, both with
an uncertainty of $\sim$ 10\% determined from planet observations
within a month from our observing runs.

The data were reduced and imaged in a standard way using MIRIAD
software \citep{Sault95}.  The uncertainty in the absolute amplitude
scale is $\sim$20 \%. The data were imaged with weights inversely
proportional to the variance in order to obtain the best signal to
noise ratio. We created a continuum map by adding together the four
windows in the upper sideband, which were found to be free of line
emission. The synthesized beam for the continuum image is 12\farcs7
$\times$ 7\farcs0 at a position angle (pa) = 4\degr, and the continuum
image has an RMS noise of  3.6 mJy~beam$^{-1}$. We easily detected
NGC\,2023\,MM\,1 with a flux density of  $\sim$ 110 mJy. However,
MM\,1 was found to be more extended than what we determined from the
JCMT SCUBA images, which appears implausible. We therefore also imaged
our phase calibrator, 0609$-$157, which should be a true point source
in the BIMA beam, and found it to have a Full Width Half Maximum
(FWHM)  of 6\arcsec $\times$ 3\arcsec, i.e. the beam is broadened by
$\sim$ 1\farcs5 -- 2\arcsec. Adding 2\farcs5 in quadrature to the
theoretical synthesized beam, we find MM\,1 to be unresolved. Since
0609$-$157 was observed only every thirty minutes, it is impossible to
judge what the true synthesized beam size was for our observations. It
is very likely that if we re-observed MM\,1 in perfect sky conditions
with the same spatial resolution, we would find it at most marginally
resolved. In order to check whether the phase errors were so large
that we would lose photons out of the main beam, we also determined
the integrated flux over the central 1\arcmin $\times$1\arcmin\
area centered on MM\,1. This gave the same flux density as we obtained
from fitting a Gaussian to MM\,1.  We therefore believe that we
retrieved all the flux, i.e. the error in flux density is dominated by
the uncertainty of the assumed flux density for 3C~273 and 0609$-$157,
both of which are variable.

We easily detected NH$_2$D 1$_{11}  \to  1_{01}$, but not SO, nor any
other molecular transition in our data set. The RMS in an individual
channel, i.e. for a velocity resolution of 0.34 km~s$^{-1}$, is $\sim$
140 mJy~beam$^{-1}$.

\subsection{SCUBA observations}

\subsubsection{SCUBA Scan maps}

We obtained several large scan maps (8\arcmin{} $\times$ 8\arcmin{})
of the NGC\,2023 region with the Submillimeter Common User Bolometer
Array (SCUBA) \citep{Holland99} on JCMT during SCUBA commissioning
time on four nights in December,1997.  SCUBA has 37 bolometers in the
long (850 $\mu$m) and 91 in the short (450 $\mu$m) wavelength array
separated by approximately two beam widths and arranged in a hexagonal
pattern.  Both arrays can be used simultaneously. 

For these observations we used the 850 $\mu$m and 450 $\mu$m filter
combination in traditional scan--map mode, i.e. with continuous
scanning in Azimuth with a scan rate of 32\arcsec~sec$^{-1}$ while
chopping in Azimuth with a chop throw of 60\arcsec. This mode gives
much flatter baselines than the more commonly used ``Emerson2'' scan
maps.  All four nights (Dec 11, 17, 18 \& 20) had excellent
submillimeter sky conditions with a precipitable water vapor level,
PWV $\leq$ 1~mm. We did a pointing check on HL~Tau or CRL\,618 before
and after each map, and corrected for any pointing drift in the post
processing of the data. During the first night, Dec 11, we had large
pointing drifts ($>$ 4\arcsec{})  during the scan map, which resulted
in smearing of point-like sources in the map, and we decided we could
not use it. In total we acquired 5 maps (12 integrations), half of
which were scanned roughly in right ascension (i.e with the source
near transit) and the rest of the maps were obtained with a roughly
orthogonal scan direction, which was achieved by observing the source
while it was rising.  HL~Tau was used as our primary calibrator,
although calibration was additionally checked using Uranus, CRL\,618
and IRC+10216. The HPBW (Half Power Beam Width) of SCUBA in scan map
mode with linear regridding to a 1\arcsec\ grid was measured to be
$\sim$ 8\farcs0 at 450 $\mu$m and $\sim$ 14\farcs8 at 850 $\mu$m from
scan maps of Uranus. 

The maps were reduced in a standard way using the SURF reduction
package \citep{Jenness99,Sandell01}.  After reducing each 850 $\mu$m
map separately, the maps were coadded and regridded onto equatorial
coordinates. This map was used to determine the position for NGC\,2023
MM\,1. We then used this position and corrected the individual maps
for residual pointing errors (shift and add) and perfomed a new coadd,
where we also weighted each map accounting for the rms-noise in each
map. The final positional accuracy is estimated to be $\leq$ 1\farcs5
based on comparison of the positions of MM\,1 (this paper) and MM\,3
\citep{wyrowski2000} deduced from BIMA observations.  The rms noise
level in the final  maps  is 45 mJy beam$^{-1}$ for the the 850 $\mu$m
map, and $\sim$ 0.45 Jy beam$^{-1}$ for the 450 $\mu$m map,

\subsubsection{ SCUBA jiggle maps}

We also took several long integration jiggle maps of the NGC\,2023
MM\,1/MM\,2 field on Dec 11 1997. These maps were reduced separately
and added to earlier jiggle maps discussed in \citet{sandell1999}. A
subset of the 850 $\mu$m maps (only those which had pointing
observations immediately before or after the map) were used to
determine the position of MM\,1. All maps were then corrected to this
position and coadded by accounting for the rms-noise in each
individual map. The jiggle maps are likely to have lower astrometric
accuracy than the scan maps, due to the limited number of pointing
checks (Table~\ref{tab_srctab}) we did for the scan maps. The rms
noise level in the jiggle maps is 30 mJy~ beam$^{-1}$ at 850 $\mu$m
and 0.23~Jy beam$^{-1}$ at 450~\micron.  The calibration accuracy is
estimated to be better than 10\% at 850 $\mu$m and about 15\% at 450
$\mu$m for both  the jiggle and the scan maps.  The flux densities
deduced for MM\,1 and MM\,2, which were observed in both modes, agree
exceptionally well, see Table~\ref{tab_srctab}.

All our maps were converted to FITS-files and exported to MIRIAD
\citep{Sault95} for further analysis. In order to correct for the
error lobe contribution, especially at 450 $\mu$m, we have deconvolved
all the maps using CLEAN and a circular model beam deduced from
observations of Uranus. For the scan maps we did the deconvolution for
only the three sub-fields where we have strong dust emission:  the
MM\,1/MM\,2 region, NGC\,2023 and MM\,3 and MM\,4,  and the MM\,5
region. The results of this analysis are discussed later in the paper.

\subsection{Ancillary molecular line data}
\label{Ancillary}

We also have access to large $^{12}$CO and $^{13}$CO $J = 2 \to 1$
scan maps obtained as bad weather backup programme on JCMT between
1995 and 1998.  The maps were obtained by scanning in RA or Dec with
sampling every 5\arcsec\ in the scan direction and stepping by
10\arcsec\ in the orthogonal direction. Since the maps were obtained
in wet weather conditions and over a long time period, each sub-map
was made to overlap with previous maps to ensure that the calibration
is uniform over the whole map. Most maps were repeated several times,
resulting in integration times per spectral data points between 5--30
seconds.  The $^{12}$CO and $^{13}$CO observations cover an area from
south of the Horsehead nebula to about the southern part of NGC\,2023,
and target the L\,1630 cloud boundary towards the IC\,434 ionization
front. To the east these observations stop just short of NGC\,2023.
The  $^{12}$CO map is $\sim$ 11\arcmin\ $\times$ 14\arcmin\  ($\alpha
\times \delta$) and contains $\sim$ 12,000 spectra with a velocity
resolution of 0.10 km~s$^{-1}$. The noise level of  the map is on the
average $\sim$ 0.15 K~km~s$^{-1}$, although there are areas where the
noise level can be five times higher. The $^{13}$CO map covers a
slightly smaller area, $\sim$ 9\arcmin\--10\arcmin\ $\times$
14\arcmin, has $\sim$ 9,600 spectra and a noise level similar to the
$^{12}$CO map. 

We have also retrieved, reduced,  and analyzed the  $^{13}$CO  and
C$^{18}$O $J = 2 \to 1$ maps from the JCMT raw data
archive\footnote{The JCMT Archive project is a collaboration between
the Canadian Astronomy Data Centre (CADC), Victoria and the James
Clerk Maxwell Telescope (JCMT), Hilo.} discussed in
\citet{johnstone2006}, as well as a smaller C$^{17}$O $J = 2 \to 1$
map of the MM\,1/MM\,2 cloud core. 

Any comprehensive analysis of these data sets are well beyond the
scope of this paper, but the large CO and $^{13}$CO maps have been
very valuable to look for outflow activity or temperature enhancements
in the vicinity of the sub-millimeter sources. Additional details of
these data set are  therefore given, at those places in the paper
where we make use of them in this study.

\subsection{Spitzer IRAC and MIPS (24~\micron) images}

We extracted observations from the Spitzer Space Observatory archive
(Program ID 43: an IRAC Survey of the L1630 and L1641 (Orion)
Molecular Clouds -- Fazio et al.).  The IRAC data were obtained
in the 12~sec High Dynamic Range (HDR) mode using four different
AORs (Astronomical Observation Requests), all with overlap. The HDR
mode takes pairs of images with 0.6 and 12~sec  frame times at each
position in all four bands. Each star in the field is therefore
covered with at least eight image frames. We have processed both short
(0.6~sec) and long (12~sec) integration basic calibrated data (BCD)
frames in each channel using the Artifact mitigation software
developed by Sean Carey and created mosaics using MOPEX. We used
APEX to identify and extract photometry of all point sources in both
the short and long integration images of the 3.6 and 4.5~\micron\
bands, while we only used the short integration images for the 5.8 and
8.0~\micron\ band.

We also extracted  MIPS 24~\micron\ observations for this region
from the Spitzer Space Observatory archive (Program ID 47:  A MIPS
Survey of the Orion L1641 and L1630 Molecular Clouds -- Fazio et al.).
The MIPS 70~\micron\ image was severely saturated and could not be
used, but the MIPS  24~\micron\ data were of good quality with only
two stars in the southern part of NGC\,2023 partially saturated.
The MIPS dataset was flatfielded, overlapped and  mosaiced using
MOPEX.

\section{Large scale overview of the NGC2023}

\begin{figure}[ht]
\begin{center}
\resizebox{\hsize}{!}{\includegraphics[angle=0,width=8.0cm,angle=0]{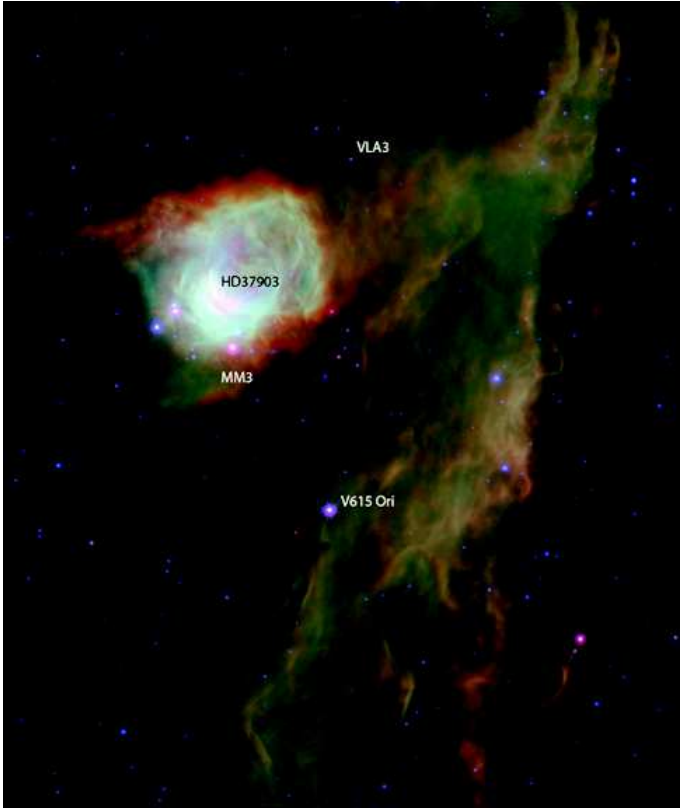}}
\caption{False-color {\em Spitzer} IRAC and MIPS image of NGC~2023. Blue:
4.5~\micron; green: 8.0~\micron; red: 24.0~\micron. The image is
centered at R.A. = 05$^h$41$^m$23$^s$ and Dec. =
-02\arcdeg19\arcmin09\arcsec, and extends over 22\arcmin$\times$26\arcmin\
($\alpha\times\delta$). N is up and E is left.
\label{fig_ch24_24um}}
\end{center}
\end{figure}

A false-color image of the NGC~2023 region is shown in
Fig.~\ref{fig_ch24_24um} (blue: 4.5~\micron; green: 8.0~\micron; red:
24.0~\micron). 
Figure~\ref{fig_n2023l_850} shows the 850~\micron\ SCUBA image
overlaid as a contour map on the Spitzer IRAC 8~\micron\ short
integration image.  Also marked in Fig.~\ref{fig_n2023l_850} are the
locations of all stars identified as young stellar objects based on
color-color diagrams incorporating 2MASS, IRAC and MIPS photometry and
the free-free emission sources VLA~1--3 identified by
\citet{reipurth2004}.  The mid-infrared and sub-millimeter images
provide a broad spectrum overview of the stellar content and dust
emission from the region. 


The emission in the mid-infrared is dominated by the reflection nebula
NGC~2023, which is illuminated by the early B-type star HD~37903.  The
images show an infrared dark dust lane at a PA of -20\arcdeg, lying
$\sim$ 7\arcmin\ inside the  Photon Dominated Region (PDR) boundary
between IC\,434 and the L1630 dark cloud. To the south-west lies the
famous Horsehead nebula, which still can be recognized in the mid-IR,
although here we only see the surface layers (PDRs) outlining the
nebula. There are a few strong mid infrared sources, but overall the
images are characterized by extended emission from the PDR. This makes
identification of discrete sources and photometry in the mid-infrared
difficult.  The physical appearance of the region in the
mid-infrared differs quite significantly from the structures seen in
low resolution IRAS and far-infrared data \citep{mookerjea2000}, the
moderate resolution 1.3~mm data \citep{launhardt1996} and the sub-mm
SCUBA data presented here.  In the far-infrared wavelengths the
emission is from the HD~37903 and the bright sources directly to the
south of it.  Owing to their poor angular resolution, the far-infrared
observations detect only a single blob of emission midway between the
dust emission peak MM~3 and HD~37903. At still longer wavelengths,
dust emission from the dark clouds dominates, with only the PDR,
illuminated by HD~37903, being marginally visible at 850~\micron.


In the SCUBA images (Fig.~\ref{fig_n2023l_850}) we detect emission
from the infrared-dark dust ridge. This dust ridge is visible in the
more extended (and deeper)  850 $\mu$m SCUBA image presented by
\citet{johnstone2006}, which shows that the dust ridge extends past
the Horsehead in the south.  (In the IRAC images the dust lane is much
narrower with very narrow ``wiggly'' filaments.) We also see strong
emission from the molecular cloud  and PDR ridge to the south and
south east of NGC\,2023 and a fainter dust lane connected to the more
prominent north-south dust ridge.  We securely identify five
sub-millimeter continuum sources named MM\,1 to MM\,5, detected at
both 450 and 850~\micron, all of which appear to be in an early stage
of evolution.  Four of the five sources detected by us were also
detected by \citet{johnstone2006} and \citet{nutter2007} using SCUBA
observations.  The sub-millimeter sources MM\,1 and MM\,2 are located
in the northern part of dust ridge.  The only sub-millimeter source
detected in the mid-infrared is MM\,3, the bright source directly to
the south of the NGC\,2023.  V615~Ori, a heavily reddened T Tauri star
lying to the north of the Horsehead nebula,  though most likely
surrounded by an accretion disk, is not detected in the
sub-millimeter.  However we detect an outflow in the \twCO\ map mostly
to the south-east of V615~Ori with relatively low velocities.

\section{Results }

\subsection{Submillimeter sources in the NGC2023 region}

\begin{figure}[t]
\begin{center}
\resizebox{\hsize}{!}{\includegraphics[angle=0,width=8.0cm,angle=0]{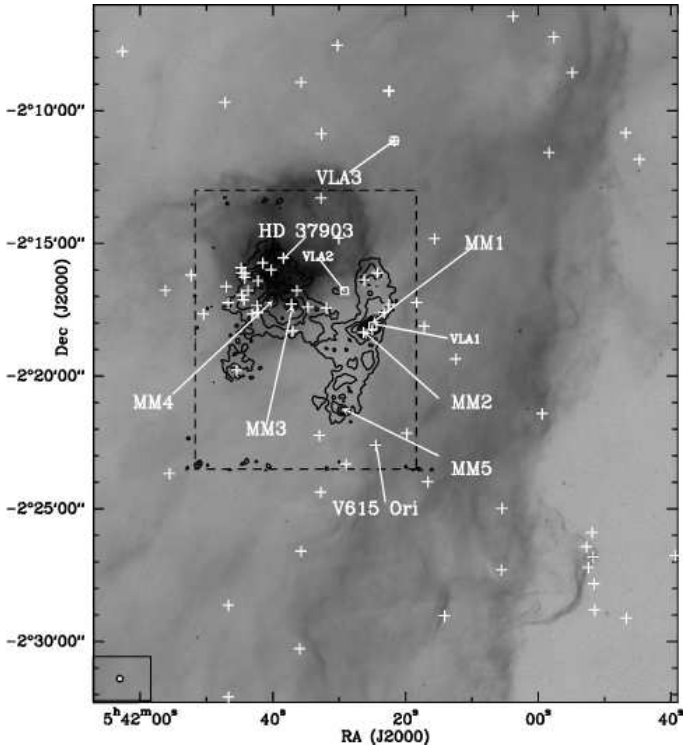}}
\caption{SCUBA 850~\micron\ contours of the entire mapped region near
NGC~2023 overlaid on the Spitzer IRAC 8~\micron\ map image.  The peak
850~\micron\ flux is 3.23~Jy/beam and the contours are at 0.22, 0.5,
0.75, 1.1, 1.5, 2, 2.5 and 3~Jy/beam. The beam at 850~\micron\ is shown
at the left bottom corner. The crosses indicate the location of the
sources identified as pre-main sequence  objects in
Fig.~\ref{fig_iracmipscol}, as well as the sources presented in
Table~\ref{tab_nirpms} and the open squares denote the location of the
free-free emission sources VLA\,1--3 \citep{reipurth2004}.
\label{fig_n2023l_850}}
\end{center}
\end{figure}

We have identified a total of five sub-millimeter continuum sources in
the region mapped with SCUBA. These are the three sources MM\,1, MM\,2
and MM\,5, located in the ridge (extended along north-south) lying to
the west of NGC~2023 and the sources MM\,3 and MM\,4 located directly
to the south and the south-east of the NGC~2023 nebula.  In order to
derive the position and sizes of the sub-millimeter sources we fitted
a two component elliptical Gaussian using the task IMFIT in MIRIAD,
one for the sub-millimeter source, and the other for the surrounding
cloud.  The fit to the broader component is mainly to provide a good
subtraction of the extended emission, and is not to estimate the flux
density of the surrounding cloud.  For all the sub-millimeter sources
the flux densities at 450 and 850~\micron\ along with positions
derived from the 850~\micron\ image and source sizes are presented in
Table~\ref{tab_srctab}.

\begin{table}[ht]
\begin{center}
\caption{Results for submillimeter sources  mapped with SCUBA
\label{tab_srctab}}
{\scriptsize
\begin{tabular}{cccccc}
\hline
\hline
\multicolumn{1}{c}{Source}& 
\multicolumn{1}{c}{$\alpha_{2000}$} & 
\multicolumn{1}{c}{$\delta_{2000}$} & 
\multicolumn{1}{c}{$S_{850}$} & 
\multicolumn{1}{c}{$S_{450}$} &
\multicolumn{1}{c}{$\theta$} \\
\multicolumn{1}{c}{}& 
\multicolumn{1}{c}{} & 
\multicolumn{1}{c}{} & 
\multicolumn{1}{c}{mJy} & 
\multicolumn{1}{c}{mJy} &
\multicolumn{1}{c}{\arcsec$\times$\arcsec} \\
 \hline
MM~1$^a$ & 5:41:24.921 & -2:18:08.7 & 3.2 $\pm$ 0.3 & 11.8 $\pm$ 1.8  & 6.5  $\times$ 5.5$^b$\\
MM~1 & 5:41:24.941 & -2:18:07.9 & 3.2 $\pm$ 0.4 & 12.1 $\pm$ 2.1 &  6.7 $\times$ 5.8\\
MM~2$^a$ & 5:41:26.331 & -2:18:18.6 & 1.5 $\pm$ 0.3 &  7.9 $\pm$ 1.3 & 15.3 $\times$ 9.5\\
MM~2 & 5:41:26.335 & -2:18:17.3 & 1.6 $\pm$ 0.5 &  9.6 $\pm$ 3.6  & 19.8 $\times$ 11.7\\
MM~3 & 5:41:37.120 & -2:17:18.1 & 1.3 $\pm$ 0.2 &  4.2 $\pm$ 0.9 & $\leq$ 6.6 $\times$ 5.6\\
MM~4 & 5:41:40.520 & -2:17:11.0 & 1.6 $\pm$ 0.4 & 11.6 $\pm$ 3.4 & 30 $\times$ 16\\
MM~5 & 5:41:29.505 & -2:21:18.3 & 0.8 $\pm$ 0.1 & 4.5 $\pm$ 0.9 &  7.1 $\times$ 5.1\\
\hline
\hline

\end{tabular} 
}
\end{center}
{\noindent
$^a$ Derived from SCUBA jiggle maps. \\ 
$^b$ Size from 450 $\mu$m 3\farcs9 $\times$ 3\arcsec{}\\}

\end{table}

The sub-millimeter emission ridge located to the west of the mapped
region extends in the north-south direction. There is no mid-IR
emission associated with this dust ridge.  Detailed study of the
northern part of this ridge using SCUBA images at 850 and 450~\micron\
images and CO $J=3\to2$ observation resulted in the identification of
MM\,1 as a class 0 protostar \citep{sandell1999}.
\citet{johnstone2006} also incorporated the SCUBA data presented here
together with additional data from the SCUBA archive. The latter data
have also been published by \citet{nutter2007}. We have completely
reprocessed the SCUBA data and have  better calibration and
astrometric accuracy compared with previously published results
\citep{sandell1999,johnstone2006,nutter2007}. \citet{launhardt1996}
detected only  the sources MM\,1, MM\,2 and MM\,3 at 1.2~mm with the
Swedish-ESO Submillimeter Telescope (SEST; beamsize of 23\arcsec) and
the IRAM 30~m telescope (beamsize =  12\arcsec).  MM\,3, located
directly to the south of the nebula is the brightest mid-infrared
(MIR) source in the
region. Surprisingly the coordinates given by \citet{launhardt1996}
for this source are significantly different from those of the 850 and
450~\micron\ sources we detect.  With the exception of MM\,1/ MM\,2
all the other sub-millimeter sources were detected at 850~\micron\ by
\citet{johnstone2006} and \citet{nutter2007}, but their positions,
sizes and deduced cloud masses differ considerably from our results.
Both of these previous studies identified only one sub-millimeter
source in the MM\,1/MM\,2 cloud core.

\citet{johnstone2006} and \citet{nutter2007} have identified 14 and 12
sources, respectively, in the region we have mapped using SCUBA.  We
can identify a few of their cores in our 850~\micron\ map, but since
they are too faint to allow secure identification at 450~\micron, we
do not list them in Table~\ref{tab_srctab}. In
Sec.~\ref{sec_disc_submm} we discuss the reasons behind the
differences between our work and these previous studies.

\subsection{Mid-infrared Photometry using IRAC and MIPS on {\em
Spitzer}
\label{sec_YSO_population}}

We carried out multiframe PSF photometry using the SSC-developed
tool APEX on all the {\em Spitzer} IRAC images and on the  MIPS
24~\micron\ images. While all  long integration  IRAC images are
considerably affected by saturation effects, the saturation of the
8~\micron\ image is particularly severe and thus could not be
corrected for by the available tools. At 8~\micron\ we therefore used
only the short integration images to identify sources and derive
photometry.  The nebulous nature of the region and strong emission
from the associated PDRs makes it extremely difficult to disentangle
the sources from the surrounding clouds and to derive accurate
photometry. Therefore a combination of automated routines and
eye-inspection was used to extract the photometry on the IRAC and MIPS
images.  For sources, which APEX failed to detect at one or several
wavelengths, we used the APEX user list option to supply the
coordinates for the source to successfully derive a PSF fit.

In all we detected over 1000 stars in the long integration  3.6 and
4.5 ~\micron\ IRAC images, while we found only 95 sources in the short
integration 8.0~\micron\ image. We obtained photometry of 44 sources
in the MIPS 24~\micron\ image. The two brightest  stars, L1630MIR-63
(MM\,3), and L1630MIR-73 \citep[star \# 219 in][]{witt1984} in the
southern part of NGC\,2023, are heavily saturated in the MIPS image
and their flux cannot be measured by conventional photometry. For
these we used the tool developed by T. Jarrett to rectify saturated
stars; this tool uses an extended PSF to fit the unsaturated wings of
the stellar light distribution. In this way we were able to recover
the integrated flux to an accuracy of $\sim$3--5\% based on the
repeatability of several measurements from the MIPS BCDs that contain
the two saturated stars.  Since we are mainly interested in young
stellar objects, we took the 95 sources detected in the short
integration IRAC 8~\micron\ images as the primary list.  We
cross-correlated this list with sources detected in other IRAC and
MIPS bands, and also with 2MASS point sources. We used the following
association radii : 1\arcsec\ for the IRAC images, 2\farcs5 for the
MIPS 24~\micron\ image and 2\arcsec\ for 2MASS data.
Table~\ref{tab_iracsrc} presents the coordinates of the 95 MIR sources
together with the 2MASS magnitudes, Spitzer-IRAC and MIPS flux
densities and a preliminary classification based on selected
color-color plots.

In order to identify the  stellar and pre main sequence (PMS) stars we
used photometry extracted from the IRAC 3.6 and 4.5 $\mu$m long
integration images, which go much deeper, have a cleaner PSF, and
appear to be less affected by nebular emission than the 8~\micron\
image. We identify 638 additional  sources, detected in both the 3.6
and 4.5~\micron\ wavebands. Of these sources 443 are found to have
2MASS point sources associated with them.  Based on a classification
scheme involving the $K_{\rm s}$--[3.6]  and [3.6]--[4.5] colors
described in Sec.~\ref{sec_ysoclass} we identify 31 of the NIR-MIR
sources to be Type II PMS stars (Table~\ref{tab_nirpms}).

We find that, with the exception of L1630MIR-35 and L1630MIR-54, all
the MIR sources detected at 8.0~\micron\ were detected in the other 3
IRAC bands as well. However only 34 MIR sources were detected in the
MIPS 24~\micron\ image and 86 MIR sources were found to have 2MASS
counterparts.  Out of the 44 sources detected at 24~\micron, 10 were
detected neither in the IRAC bands nor in the sub-millimeter images
(Table~\ref{tab_onlymips}). Six of the sources detected only in the
MIPS band only are in or near NGC\,2023. Of the MIPS-only sources
(Table~\ref{tab_onlymips}) L1630MIPS-4 is close to MM\,2,and
L1630MIPS-5 is in the vicinity of MM\,1 and (probably) the young stars
associated with the cold MM\,1 cloud. Several other MIPS sources lie
close to faint sub-millimeter peaks that we do not identify as
sub-millimeter sources in this paper (Fig.~\ref{fig_n2023l_850}).
There are several interesting sources (VLA3, Sellgren's star C,
HD~37903, and V615~Ori), which are bright in the MIR, but which were
not detected sub-millimeter; see Section~\ref{sec_onlymir} for further
details.

\subsubsection{YSO classification based on NIR-MIR colors
\label{sec_ysoclass}}

\begin{figure*}[ht]
\begin{center}
\resizebox{\hsize}{!}{\includegraphics[angle=0,width=16.0cm,angle=0]{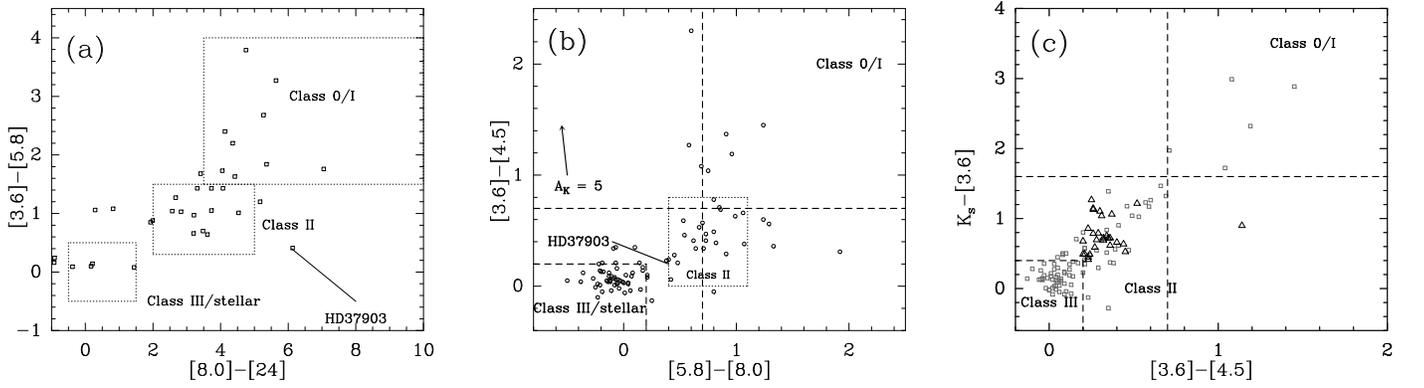}}
\caption{Color-color diagrams for all MIR and PMS-NIR sources.
Approximate classification criteria adopted from
\citet{megeath2004} and \citet{muzerolle2004} are shown respectively
for the left and the middle panel. (a) MIR color-color diagram based
on {\em Spitzer} 3-band IRAC and MIPS photometry. (b) 4 band IRAC
color-color diagram and (c) NIR-MIR color-color diagram based on 2MASS
$K_{\rm s}$ and two {\em Spitzer} IRAC bands.  Dashed lines in all
panels are taken from \citet{hartmann2005}. Reddening vector
corresponding to the extinction laws for NGC2023/2024 given by
\citet{flaherty2007} is shown in {\em (b)}. The dashed lines $K_{\rm
s}$--[3.6] = 1.6, [3.6]--[4.5] = 0.7, [4.5]--[5.8] = 0.7 and
[5.8]--[8.0] = 0.7 discriminate Class II sources from Class I/0
sources, and $K_{\rm s}$--[3.6] = 0.4, [3.6]--[4.5] = 0.2,
[4.5]--[5.8] = 0.2 and [5.8]--[8.0] = 0.2 discriminate Class III from
Class II. Squares and triangles in {\em (c)} correspond to sources
with and without detection at 8~\micron\ respectively.
\label{fig_iracmipscol}}
\end{center}
\end{figure*}

We classified the sources detected in the NGC~2023/L~1630 region using
the NIR and MIR photometric data presented above.
Figure~\ref{fig_iracmipscol} presents the color-color diagrams derived
from the 2MASS, IRAC and MIPS 24~\micron\ magnitudes of sources
detected in NGC\,2023 along with several criteria (shown as dashed
lines in Fig.~\ref{fig_iracmipscol}) used to classify them.

Based on the combined IRAC+MIPS colors ([3.6]--[5.8] vs
[8]--[24]) (see Figure~\ref{fig_iracmipscol} ({\em
left})) we find  that almost all the sources have significant infrared
excess.  Most of the sources that have been identified in the four
IRAC and MIPS bands belong to Class II.  At 24~\micron\ the
photospheric colors should be close to zero for all spectral types,
hence the [8]--[24] color is very sensitive to excesses.



For the sources detected in all four IRAC bands we find that both the
[5.8]-[8.0] and the [3.6]--[4.5] colors for these sources extend
over 1.5 mag.  (Fig.~\ref{fig_iracmipscol} {\em middle}).  Sources
with the colors of stellar photospheres are centered at
([3.6]--[4.5],[5.8]--[8.0])=(0,0) and include foreground and
background stars  as well as diskless (Class III) pre-main sequence
stars. The box outlined in Fig.~\ref{fig_iracmipscol} ({\em middle}),
defines the location of Class II objects in the plane
\citep{megeath2004,allen2004}.  These sources have colors that can be
explained by young, low-mass stars surrounded by disks.
\citet{hartmann2005} have shown from their observations of
Taurus-Auriga young stars that the color criteria
[3.6]--[4.5]$>0.7$ and [4.5]--[5.8]$>0.7$ isolate Class 0/I
protostars from Class II and Class III T Tauri stars. 

\citet{hartmann2005} have also shown that the color criteria $K_{\rm
s}$--[3.6]$>1.6$ and [3.6]--[4.5]$>0.7$ discriminate well between
Class II and Class 0/I systems. We have thus used the photometry for
the sources presented in Table~\ref{tab_iracsrc} which have 2MASS
counterparts to generate the $K_{\rm s}$--[3.6] versus [3.6]--[4.5]
color-color plot (Fig.~\ref{fig_iracmipscol} {\em right}). In the same
color-color plot we also show the 31 (out of the 443 sources detected
in the 3.6 and 4.5~\micron\ IRAC images and with 2MASS associations )
which are identified to be PMS in nature based on the color criteria
(Table~\ref{tab_nirpms}). We do not plot the entire sample of 443
sources in order to avoid crowding the figure. Owing to severe
contamination due to PAH emission at 3.6~\micron\ L1630NIR-23 and
L1630NIR-24 do not show colors typical of YSOs, but since they are
known to be H$\alpha$ emission stars, we count them as YSOs and
included them in Table~\ref{tab_nirpms}.

\setcounter{table}{2}
\begin{table*}
\begin{center}
\caption{Coordinates and flux densities of 2MASS \& IRAC 3.6/4.5 $\mu$m PMS sources in NGC~2023/L~1630.}
\label{tab_nirpms}
{\small
\begin{tabular}{lcccccccc}
\hline
\hline
Source & $\alpha_{2000}$ & $\delta_{2000}$ & $J$ & $H$ & $K_s$ &
F$_{3.6}$ & F$_{4.5}$ & Name$^{d}$\\
L1630 & & & &  & &mJy &mJy &\\
\hline
NIR-1  &  5:40:46.70 &  -2:29:07.1 & 16.36$\pm$0.09& 15.75$\pm$0.10& 15.33 $\pm$0.16&      0.40 $\pm$ 0.01&      0.36 $\pm$ 0.00 &  \\
NIR-2  &  5:40:52.70 &  -2:26:25.8 & 16.69$\pm$0.13& 15.42$\pm$0.08& 15.15 $\pm$0.14&      0.37 $\pm$ 0.01&      0.29 $\pm$ 0.00 & \\
NIR-3  &  5:40:54.87 &  -2:08:33.7 & 17.05$\pm$0.18& 15.66$\pm$0.09& 15.07 $\pm$0.13&      0.49 $\pm$ 0.01&      0.38 $\pm$ 0.01 & \\
NIR-4  &  5:40:57.64 &  -2:07:13.2 & 17.68$\pm$0.10& 16.00$\pm$0.10& 14.93 $\pm$0.14&      0.78 $\pm$ 0.01&      0.66 $\pm$ 0.01 & \\
NIR-5  &  5:40:58.36 &  -2:11:34.6 & 15.62$\pm$0.06& 14.38$\pm$0.04& 13.61 $\pm$0.05&      2.09 $\pm$ 0.02&      1.33 $\pm$ 0.01 & \\
NIR-6  &  5:40:59.39 &  -2:21:24.8 & 14.82$\pm$0.04& 14.18$\pm$0.03& 13.59 $\pm$0.05&      2.14 $\pm$ 0.02&      1.78 $\pm$ 0.01 & \\
NIR-7  &  5:41:05.54 &  -2:27:17.9 & 18.37$\pm$0.01& 16.04$\pm$0.13& 14.36 $\pm$0.07&      1.64 $\pm$ 0.01&      1.31 $\pm$ 0.01 & B33-31\\
NIR-8  &  5:41:14.12 &  -2:29:01.6 & 17.01$\pm$0.18& 15.84$\pm$0.12& 15.68 $\pm$0.26 & 0.42$\pm$0.01 & 0.234$\pm$0.01 & B33-33\\
NIR-9  &  5:41:15.61 &  -2:14:49.1 & 18.70$\pm$0.10& 16.21$\pm$0.17& 15.10 $\pm$0.15&      0.44 $\pm$ 0.01&      0.36 $\pm$ 0.01 & \\
NIR-10 &  5:41:18.30 &  -2:17:13.6 & 16.83$\pm$0.15& 15.10$\pm$0.06& 14.38 $\pm$0.08&      0.72 $\pm$ 0.01&      0.57 $\pm$ 0.01 & \\
NIR-11 &  5:41:19.79 &  -2:22:09.2 & 18.27$\pm$0.10& 16.01$\pm$0.13& 14.87 $\pm$0.13&      0.50 $\pm$ 0.01&      0.38 $\pm$ 0.01 & \\
NIR-12 &  5:41:22.48 &  -2:09:15.0 & 16.93$\pm$0.10& 15.74$\pm$0.13& 14.60 $\pm$0.11&      0.74 $\pm$ 0.01&      0.69 $\pm$ 0.01 & \\
NIR-13 &  5:41:22.49 &  -2:17:18.6 & 14.68$\pm$0.04& 12.95$\pm$0.03& 12.05 $\pm$0.02&      8.05 $\pm$ 0.03&      6.89 $\pm$ 0.03 & \\
NIR-14 &  5:41:30.06 &  -2:14:47.7 & 17.67$\pm$0.10& 15.54$\pm$0.10& 14.42 $\pm$0.08&      0.85 $\pm$ 0.01&      0.76 $\pm$ 0.01 & \\
NIR-15 &  5:41:31.93 &  -2:17:25.7 & 17.19$\pm$0.10& 16.03$\pm$0.10& 15.00 $\pm$0.17&      0.64 $\pm$ 0.01&      1.18 $\pm$ 0.01 &  HH 247\\
NIR-16 &  5:41:32.67 &  -2:10:51.9 & 16.93$\pm$0.14& 15.14$\pm$0.06& 14.46 $\pm$0.09&      0.72 $\pm$ 0.01&      0.56 $\pm$ 0.01 & \\
NIR-17 &  5:41:32.73 &  -2:13:17.5 & 17.29$\pm$0.23& 15.55$\pm$0.10& 14.47 $\pm$0.08&      0.82 $\pm$ 0.01&      0.79 $\pm$ 0.01 & \\
NIR-18 &  5:41:32.75 &  -2:24:22.0 & 18.65$\pm$0.10& 16.20$\pm$0.15& 14.74 $\pm$0.10&      1.01 $\pm$ 0.01&      0.82 $\pm$ 0.01 & \\
NIR-19 &  5:41:32.97 &  -2:22:14.1 & 17.32$\pm$0.10& 16.16$\pm$0.17& 15.17 $\pm$0.14&      0.53 $\pm$ 0.01&      0.42 $\pm$ 0.01 & \\
NIR-20 &  5:41:35.74 &  -2:08:55.8 & 17.20$\pm$0.10& 15.61$\pm$0.11& 14.69 $\pm$0.10&      0.59 $\pm$ 0.01&      0.47 $\pm$ 0.01 & \\
NIR-21 &  5:41:35.74 &  -2:26:35.6 & 18.73$\pm$0.10& 15.95$\pm$0.12& 14.67 $\pm$0.10&      0.79 $\pm$ 0.01&      0.64 $\pm$ 0.01 & \\
NIR-22 &  5:41:35.95 &  -2:30:16.3 & 18.59$\pm$0.10& 16.17$\pm$0.16& 14.80 $\pm$0.11&      0.93 $\pm$ 0.01&      0.78 $\pm$ 0.01 & \\
NIR-23$^a${} & 5:41:41.51 & -2:15:43.9 & 11.38$\pm$0.11 &   11.63$\pm$ 0.11 & 14.10$\pm$0.04 & 7.86$\pm$0.08 $^b$ & 4.00$\pm$0.04 & WB~59\\
NIR-24$^a$ &  5:41:43.09 & -2:17:39.8  & 13.03$\pm$0.03 & 12.35$\pm$0.02 & 12.11$\pm$0.03 & 4.90$\pm$0.03 & 3.21$\pm$0.02 & WB~57\\
NIR-25$^c$ &  5:41:44.22 &  -2:16:07.5 & 13.38$\pm$0.04 & 12.51$\pm$0.05& 11.93$\pm$0.04&     14.67 $\pm$0.08&     15.03$\pm$0.07 & 13/W218\\
NIR-26 &  5:41:45.48 &  -2:19:46.4 & 17.52$\pm$0.10& 16.28$\pm$0.18& 14.95$\pm$0.14&      0.85$\pm$0.01&      0.69$\pm$0.01 & \\
NIR-27 &  5:41:46.67 &  -2:28:38.1 & 16.72$\pm$0.14& 15.98$\pm$0.12& 15.24$\pm$0.17&      0.45$\pm$0.01&      0.39$\pm$0.00 & \\
NIR-28 &  5:41:46.70 &  -2:17:13.9 & 16.92$\pm$0.16& 16.20$\pm$0.18& 15.26$\pm$0.18&      0.43$\pm$0.01&      0.38$\pm$0.01 & \\
NIR-29 &  5:41:50.44 &  -2:17:39.6 & 18.20$\pm$0.10& 16.39$\pm$0.20& 14.56$\pm$0.12&      1.11$\pm$0.01&      1.01$\pm$0.01 & \\
NIR-30 &  5:41:52.31 &  -2:16:11.8 & 16.42$\pm$0.10& 14.25$\pm$0.07& 12.97$\pm$0.04&      3.46$\pm$0.02&      2.88$\pm$0.01 & \\
NIR-31 &  5:41:56.21 &  -2:16:45.9 & 16.86$\pm$0.18& 16.30$\pm$0.20& 15.15$\pm$0.10&      0.40$\pm$0.01&      0.39$\pm$0.01 & \\
\hline
\hline
\end{tabular}
}
\end{center}
{\noindent
$^a$ H$\alpha$ emission line star \citep{weaver2004}\\
$^b$ 3.6~\micron\ photometry is unreliable due to severe contamination 
from PAH emission\\
$^c$ Completely hidden in the Airy pattern of the of the exceedingly
bright star L1630MIR-72 at both 5.8 and 8.0~\micron. \\ 
$^d$  Integers correspond
to NIR sources detected by \citet{depoy1990}, 
WB stands for sources
detected by \citet{weaver2004}, W are sources detected by
\citet{witt1984} and B33 are names from \citet{bowler2009}}
\end{table*}

A summary of the classification of sources is presented in
Table~\ref{tab_iracsrc}.  For most sources it was possible to arrive
at a unique class based on all three color-color diagrams. For all
sources the class derived based on at least two color-color diagrams
match. For sources indicating two different classes from the three
color-color diagrams, we have indicated both classes.  Out of the 95
sources detected in the IRAC 8~\micron\ band 32 stars are found to be
PMS stars (Classes I, II and I/II) based on their colors. Two sources
detected only in the IRAC 5.8 and 8.0~\micron\ (L1630MIR-35 and 54) are
most likely  PMS, while an additional 14 stars marked as type II/III
may or may not be PMS stars.  The 10 MIPS-only sources
(Table~\ref{tab_onlymips}) are likely to be PMS stars, although we
cannot exclude the possibility of some of these stars being background
stars.  Use of the more sensitive long integration 3.6 and
4.5~\micron\ images in combination with 2MASS data yields an
additional 31 PMS objects (Table~\ref{tab_nirpms}). Since the long
integration 3.6 and 4.5~\micron\ images go deeper and are less
affected by extinction than the 2MASS survey, we also checked how many
of these IRAC sources are likely to be PMS stars.  Since there are no
strong PAH (polycyclic aromatic hydrocarbon) features in the 4.5
\micron\ band, and the 3.3 and 3.4~\micron\ PAH bands will be in
emission rather than absorption, the
[3.6]--[4.5] color
should be sensitive to stars with a very red continuum, i.e. stars
with infrared excess. If we use the color criterion, [3.6]-[4.5]
$\geq$ 0.2, for sources detected only in 3.6 and 4.5~\micron\ which
have no 2MASS counterparts, we find another 38 sources which are
likely to be PMS stars.  In Table~\ref{tab_iracsrc} we denote the
sources which exclusively occupy the region characteristic of the
stellar photospheres in the color-color diagram as ``S", although some
of them may indeed be young stars.

Figure~\ref{fig_n2023l_850} shows the PMS objects identified in the
region with vertical white crosses.  Thus in the NGC~2023 region we
identify 73 out of a total of 739 sources (111, if we include the very
red sources seen only in the IRAC 3.6 and 4.5 \micron\  images) to be
PMS in nature and of these five are bona fide Class I sources. Nine are
identified as Class I/II. We discuss the distribution of these YSOs in
Sec.~\ref{sec_discussion}. 

\begin{figure*}[t]
\begin{center}
\includegraphics[width=14.0cm]{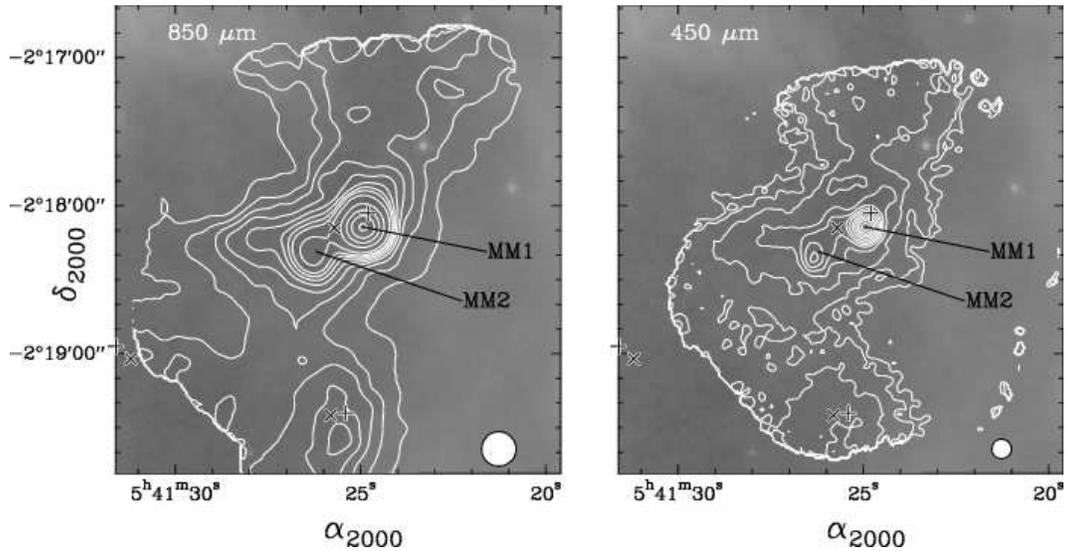}
\caption{Contours of SCUBA 850 and 450~\micron\ continuum jiggle maps of
the field surrounding NGC\,2023 MM\,1 and MM\,2 overlaid on the grayscale
image of Spitzer IRAC 8~\micron. The peak intensities at 850 and
450~\micron\ are 3.08 and 9.6~Jy/beam, respectively. The contour levels
for the 450~\micron\ map are at 0.3, 0.5, 1, 1.75, 2.5, 3.25, 4, 5, 6,
7, 8  and 9~Jy/beam.  The contour levels for the 450~\micron\ map are at 0.1,
0.2, 0.3, 0.4, 0.5, 0.6, 0.7, 0.8, 1, 1.5, 2.0, 2.5 and 3.0~Jy/beam. 
The beam sizes at 450 and 850~\micron\ are shown at the right bottom corner
of each panel. ``+"  and ``x" signs mark the sub-mm sources detected
in this region by \citet{johnstone2006} and \citet{nutter2007}.
\label{fig_n2023MM1}}
\end{center}
\end{figure*}

\begin{table}[ht]
\begin{center}
\caption{Sources detected only in the MIPS 24~\micron\ band
\label{tab_onlymips}}
{\small
\begin{tabular}{lllr}
\hline
\hline
Source & $\alpha_{2000}$ & $\delta_{2000}$ & F$_{24}$\\
L1630& & & mJy\\
\hline
MIPS-1  &  5:40:51.48 &  -2:28:48.5 &       2.9$\pm$0.2 \\ 
MIPS-2  &  5:40:51.57 &  -2:27:49.2 &       6.0$\pm$0.2 \\ 
MIPS-3  &  5:40:51.84 &  -2:25:53.5 &       2.7$\pm$0.2 \\ 
MIPS-4  &  5:41:16.64 &  -2:23:58.8 &      34.7$\pm$0.2 \\ 
MIPS-5  &  5:41:25.02 &  -2:18:15.3 &       2.0$\pm$0.2 \\ 
MIPS-6  &  5:41:26.35 &  -2:18:21.1 &      11.2$\pm$0.2 \\ 
MIPS-7  &  5:41:37.03 &  -2:18:17.8 &      44.1$\pm$0.2 \\ 
MIPS-8  &  5:41:43.75 &  -2:16:46.8 &      37.4$\pm$0.2 \\ 
MIPS-9  &  5:41:44.36 &  -2:17:07.9 &      55.9$\pm$0.2 \\ 
MIPS-10  &  5:41:47.20 &  -2:09:40.8 &       6.0$\pm$0.1 \\ 
\hline
\hline
\end{tabular}
}
\end{center}
\end{table}

\section{The NGC2023 MM\,1/MM\,2 region
\label{results_mm1_mm2}}

As mentioned earlier the SCUBA data for the NGC~2023 MM\,1/MM\,2
region is of improved quality compared to the data presented by
\citet{sandell1999}. We determine the position of MM\,1 more
accurately with BIMA and we resolve MM\,2 in both SCUBA bands. We do
not detect MM\,1 with either 2MASS, IRAC or MIPS. The flux densities
for MM\,1 and MM\,2 at 450 and 850~\micron\ (Table~\ref{tab_srctab})
are within 10\% of the flux densities measured by \citet{sandell1999}.

\citet{sandell1999} showed that the sub-millimeter source MM\,1
satisfies all the criteria for a low mass class 0 protostar. It is not
detected in the near-infrared. It powers a jet-like outflow with
extreme velocity ($\Delta$v $>$ 200~\kms) detected in the CO $J = 3
\to 2$ \citep{sandell1999}.  Its outflow efficiency, defined as
the ratio of the momentum flux due to the mechanical force to the
radiative momentum flux ($F_{\rm CO}$/$F_{\rm rad}$), is $>1000$,
a value typically seen in class 0 sources.  A large fraction of its
luminosity is emitted in the sub-millimeter $L_{\rm submm}/L_{\rm bol}
> 10^{-3}$ with $L_{\rm bol}<10$~\lsun. They found the protostellar
disk to be extended in the sub-millimeter with a size $\sim 5$\arcsec,
i.e., diameter $\sim 2000$~AU and a mass of 2--4~\msun.
\citeauthor{sandell1999} found the  dust emission to be significantly
optically thick at 850~\micron.

\subsection{BIMA results}

\begin{figure}[ht]
\begin{center}
\resizebox{\hsize}{!}{\includegraphics[width=8.0cm]{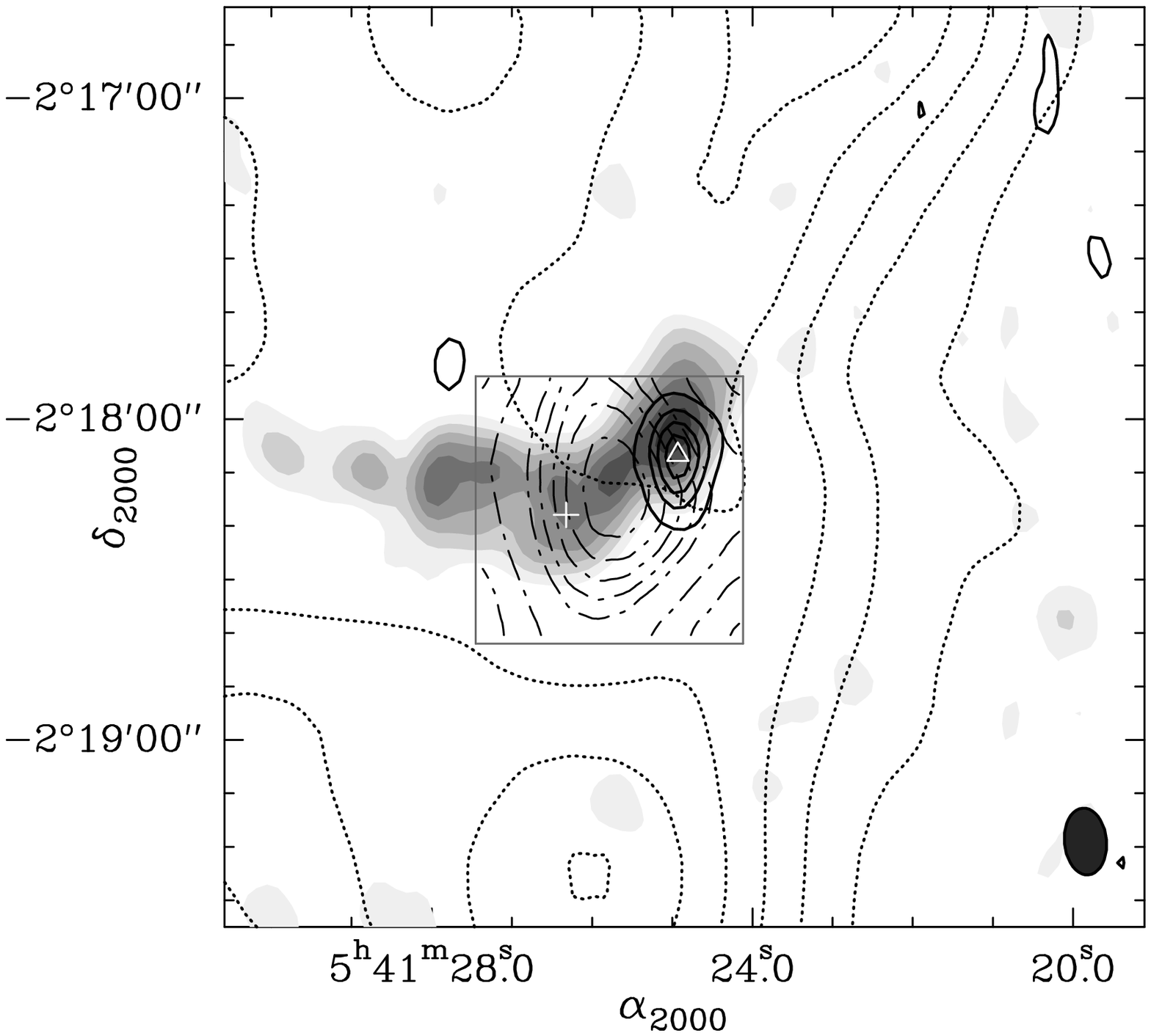}}
\caption{The \nhtwod\ emission from the MM1/MM2 region is plotted in
greyscale overlaid with contours of  \CeiO\ $J = 2 \to 1$
(dotted), \COseventeen\ $J = 2 \to 1$ (dash-dot) and 3.5~mm continuum
(solid black). The extent of the \COseventeen\ map, which only covers a
portion of the image is shown with a grey square.  The \nhtwod\
grey levels are obtained by integrating over the main component and  are
drawn at 0.2, 0.4, 0.6, 0.8, 1, 1.2, 1.4 and 1.6~K~km~s$^{-1}$. The 3.5~mm
continuum emission  is centered on MM\,1 with contours drawn at 0.010,
0.025, 0.04, and 0.055~Jy/beam. The BIMA beam is shown in the bottom
right corner. The  \CeiO\ and \COseventeen\ $J = 2 \to 1$ emission is
integrated over a 1~\kms\ wide window centered at 9.4~\kms, 
the systemic velocity of the cloud.  The contours for
\CeiO\ are from 1\% to 100\% in steps of 10\% of the peak value,
7.1~K~km~s$^{-1}$. The contours for \COseventeen\ are at 0.7, 0.75,
0.8, 0.85, 0.9, 0.95, 0.97, 0.99, 1.01, 1.03 and 1.05~K~km~s$^{-1}$. The
triangle marks the position of the 3.6~cm free-free emission source
VLA\,1 (coinciding with MM\,1), while the white cross marks the
position of the SCUBA continuum source MM\,2. 
\label{fig_n2023bima}}
\end{center}
\end{figure}

Figure~\ref{fig_n2023bima} shows the contours of the 3.5~mm continuum
and intensity of \nhtwod\ integrated over the main component observed
with BIMA, overlaid on the 450~\micron\ SCUBA image. We also overplot
the contours of integrated intensities of \COseventeen\ and \CeiO\ $J
= 2 \to 1$ in Fig.~\ref{fig_n2023bima} for reference.  The
sub-millimeter source MM\,1 is readily detected at 3.5~mm, but not
MM\,2. \nhtwod\ emission shows a distribution similar to the continuum
emission and is detected from both MM\,1 and MM\,2. Strong emission
from the deuterated molecule could be a signature of the youth of the
system.  The peak of the \nhtwod\  is offset to the north by
$\sim$ 1\farcs5 in declination with respect to the continuum peak at
MM~1 in the BIMA image.  Since the relative positional uncertainty
between the spectroscopic and continuum observations done with BIMA
simultaneously is negligible, the detected offset is significant.
\citet{roueff2005} also found that the \nhtwod\  emission does not
seem to peak at the positions of the embedded protostars, but instead
at offset positions, possibly due to interaction between the ambient
cloud and the outflow.  A plausible explanation could be that
\nhtwod\ gets destroyed once disks form, so that only in the
pre-stellar phase the emission is peaked on the protostar. For
somewhat more evolved protostars it no longer peaks at the center of
the source.  Further, at moderate velocities the outflow could sputter
\nhtwod\ molecules off dust grains, resulting in an enhancement of
\nhtwod. This of course requires that the outflow velocities are not
high enough to destroy the grains completely, but are moderate as
found in the shearing layer between the outflow and the ambient
cloud.

Although MM\,1 appears extended in our BIMA observations
(Section~\ref{BIMA_obs}), we do not know whether we resolved the
continuum emission or not, because of the marginal weather conditions
during our observations. The poor phase stability may also affect the
position of the continuum emission; $\alpha$(2000.0) =
05$^h$~41$^m$~24\farcs929, $\delta$(2000.0) =
$-$02\degr{}18\arcmin{}~07\farcs2, with an uncertainty of 0\farcs8,
estimated from dividing up the data into subsets and determining the
position of MM\,1 in each data set. This position agrees within errors
with the faint free-free emission source NGC~2023\,VLA\,1
\citep{reipurth2004}, offset by $-$0\farcs0,+0\farcs5 in RA and Dec,
respectively. Since the free-free emission is only 0.14 mJy at 3.6 cm,
the contribution from free-free emission is completely negligible at
3.5~mm, i.e.  all the continuum emission at 3.5~mm is due to dust.

\subsection{Is \nhtwod\ tracing the protostellar disk of MM~1?}

\begin{figure*}[t]
\begin{center}
\includegraphics{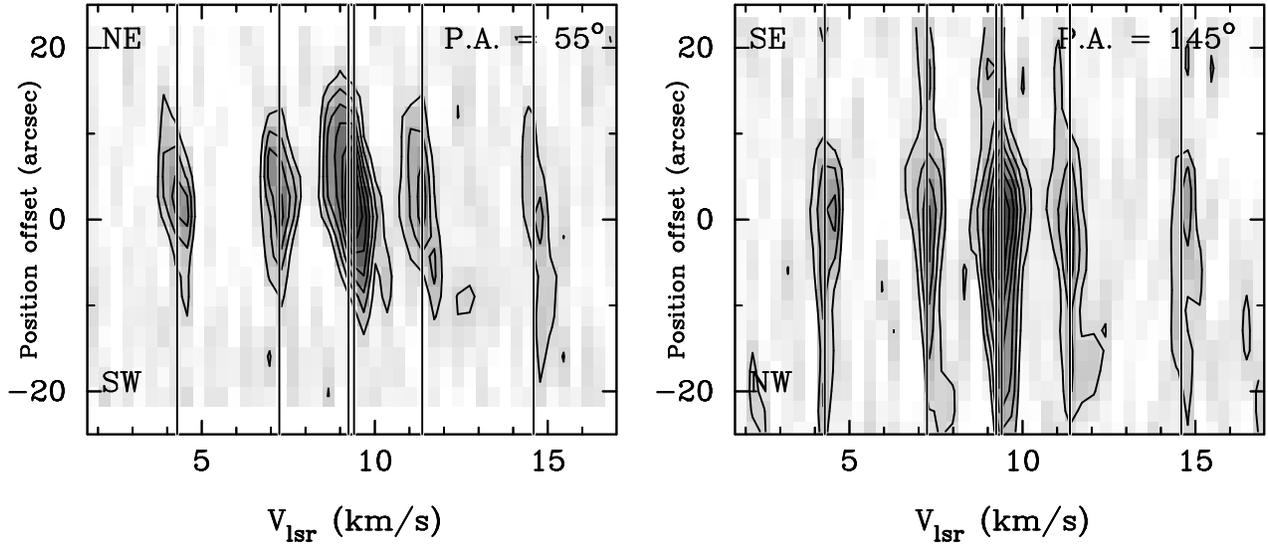}
\caption{Position velocity diagram of the \nhtwod\ emission along (a)
the assumed disk orientation (PA=55\arcdeg) and (b) along the outflow
(PA=145\arcdeg) centered at MM\,1.The vertical lines denote the six 
hyperfine spectral
lines with \vlsr=9.4~\kms\ for the main component derived from Gaussian 
profile fitting. 
\label{fig_nh2dpv}}
\end{center}
\end{figure*}

Based on CO J=3$\rightarrow$2 observations \citet{sandell1999} concluded
that NGC~2023~MM\,1 drives a strong outflow at a position angle of
145\arcdeg (measured from north to east). To explore the presence or
detection of a disk from our \nhtwod\ data we assume that the disk is
oriented at right angles to the outflow with PA = 55\arcdeg.
Figure~\ref{fig_nh2dpv} shows the position-velocity (PV) diagram of
the \nhtwod\ emission along the assumed disk as well as the outflow.
We see a distinct velocity gradient along the disk, suggesting that
the protostellar disk or the circumstellar material is rotating. The
emission is slightly redshifted towards northeast and blueshifted to
the southwest with marginal asymmetry relative to the position of the
continuum source which is at an offset of (+1\farcs2,+1\farcs8) relative
to the phase center for the BIMA maps.  

\begin{figure}[ht]
\begin{center}
\includegraphics[angle=0,width=8.0cm,angle=0]{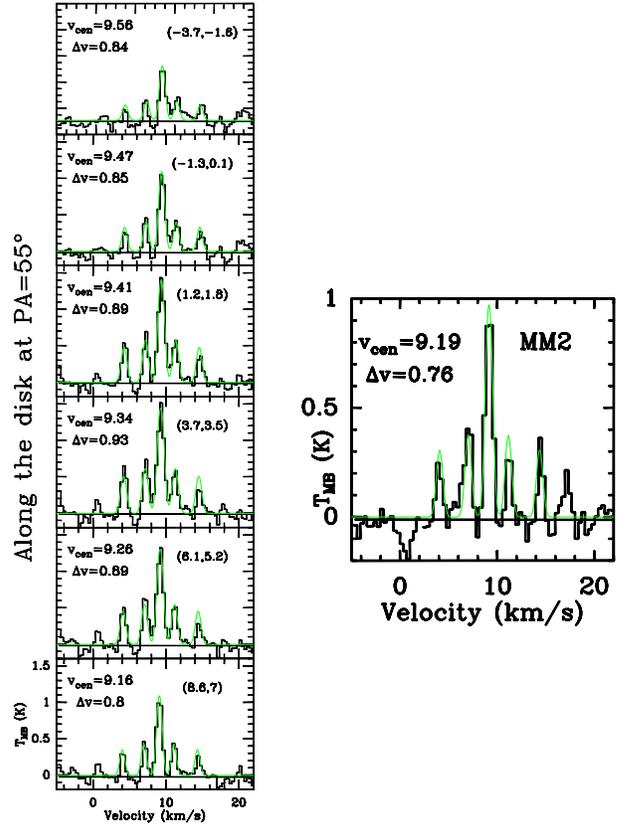}
\caption{\nhtwod\ spectra along the disk (Left: PA = 55\arcdeg) at MM\,1 
and the
spectrum at the position of MM~2 (Right).  The green curve in each panel
shows the results of the hyperfine structure fitting obtained using the
HFS method in GILDAS/CLASS. In each panel the velocity of the main
component of the hyperfine lines as well as the fitted Gaussian widths
are given.
\label{fig_nh2dspec}}
\end{center}
\end{figure}

Figure~\ref{fig_nh2dspec} shows spectra at selected positions along
the disk with MM\,1 at the center and at the position of MM\,2. 
The spectra were analyzed using CLASS method HFS to fit the six
hyperfine lines. The HFS method of CLASS is a part of the
GILDAS\footnote{http://www.iram.fr/IRAMFR/GILDAS}
software package which calculates the total optical depth $\tau$, the
average linewidth of  lines  and the radiation temperature of the
multiplet, assuming LTE.  Frequencies and line strengths were adopted
from \citet{tine2000}. The derived velocities in the LSR frame,
\vlsr, for the main component, the optical depths and linewidths
resulting from the fit are presented in Table~\ref{tab_hfs}. Based on
the fits we find that there is distinct shift in \vlsr\ along the
disk, the shift is of the same magnitude as the velocity resolution
for the BIMA observations. Although the velocity resolution of the
BIMA observations is 0.334 km/s, based on the assumption that the
lines are Gaussian it is possible to determine the center of the line
to at least an accuracy of better than a tenth of this value owing to
the improved accuracy of the fits due to the hyperfine lines.  We find
that the velocity changes by 0.13~\kms\ over a lengthscale of $\pm
3$\arcsec. Most significantly the velocity change per unit length,
closer to the center of the disk is similar throughout the length of
the disk.  If the velocity gradient was due to Keplerian motion, a
smaller gradient would have been observed closer to the star.  This
implies that the rotating disk which probably lies hidden inside is
not seen in \nhtwod, the emission appears to be primarily due to the
large scale infalling envelope rotating approximately as a rigid body.

The narrow line widths seen in \nhtwod\ also support this non-detection.
If \nhtwod\ were probing a rotating disk, the emission from the disk in
our relatively broad beam, should by itself broaden the line. \nhtwod\
is quickly destroyed in shocks and in warm gas, so that it may be absent
in the accretion disk and what we see is the pristine cold gas in the
surrounding, slowly rotating infalling envelope.  We note that
although \nhtwod\ could exist in the dense cold mid-plane of the disk,
the size of the disk and the low temperatures of the mid-plane would
make detection of the emission from this region extremely difficult.


We find that in comparison to the direction along the disk there is hardly
any velocity gradient along the direction of the outflow
(Fig.~\ref{fig_nh2dspec}). This implies that the spread in velocity
seen in the P-V diagram (Fig.~\ref{fig_nh2dpv}) is primarily due to
the broadening of the spectral lines by the outflow and the difference
in velocities of MM~1 and MM~2 (Tab.~\ref{tab_hfs}). 

\subsection{Mass of MM\,1 and MM\,2 from greybody SED fits
\label{sec_mm1_greyb}}

\begin{figure}[ht]
\begin{center}
\resizebox{\hsize}{!}{\includegraphics{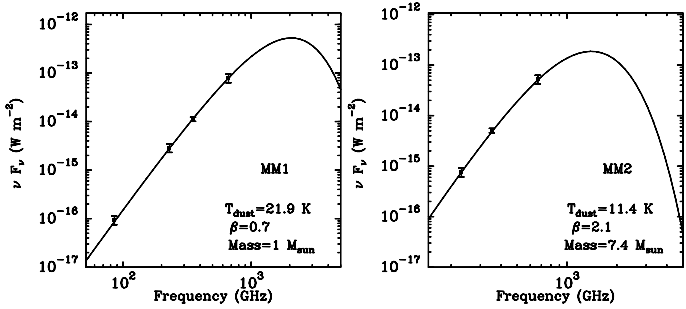}}
\caption{Single temperature greybody fits to the SEDs of NGC~2023 MM~1
({\em left}) and MM~2. The fitted dust temperature, \tdust, the
dust emissivity index ($\beta$) and the total mass, M$_{\rm tot}$ for
the best fit model are given.
\label{fig_greyfit}}
\end{center}
\end{figure}

Figure~\ref{fig_greyfit} shows the greybody models fitted to the
observed SEDs of NGC~2023-MM\,1 and MM\,2. Based on the observed sizes
($\theta_s$) of the sources, for the greybody fitting we have assumed
$\theta_s$ to be 5\arcsec\ and 10\arcsec\ for MM\,1 and MM\,2
respectively. The dust temperatures (\tdust), dust emissivity indices
($\beta$) and masses calculated for the best fit model are presented
in Fig.~\ref{fig_greyfit}. The total dust and mass gas, $M_{\rm tot}$
has been derived from the fitted \tdust\ and $\beta$ using

\begin{equation}          
M_{tot} = 1.88 \times 10^{-2} \biggl({1200\over\nu}\biggr)^{3+\beta}S_\nu
(e^{0.048\nu/T_d}-1)d^2,
\end{equation}

where d is the distance in $kpc$, S$_\nu$ is the total flux (in
Jy) at frequency $\nu$ and  M$_{\rm tot}$ is given in $[$\msun$]$.  In
this equation we have adopted  a gas-to-dust ratio of 100 and standard
``Hildebrand'' mass opacity, ${\rm \kappa_o}$, defined at 250~$\mu$m
(1200 GHz), i.e.  $\kappa_{\rm 1200GHz}$ = 0.1 cm$^{2}$g$^{-1}$
\citep{hildebrand1983}.

The $\beta$-index derived for  MM\,1, 0.7, is much smaller than what
was found by  \citet{sandell1999}. However, since their fit was
dominated by single beam sub-millimeter photometry, which includes emission
from the surrounding cloud, their  $\beta$-index was too high, and
their fit underestimated the SED at long wavelengths. We subtract out
the emission from the surrounding cloud, and since we now also have a
flux density measurement at 3.5~mm, we obtain a very good fit.
A $\beta$ of 0.7, though apparently low for a Class 0
protostar \citep{froebrich2005},  is not unusual. HH\,24~MMS,
another Class 0 source in the L\,1630 dark cloud complex, has a very
similar $\beta$-index \citep{chandler1995,kang2008}.  We estimate the
total mass of MM\,1 to be 1.04~\msun, lower than the mass estimate of
\citet{sandell1999}, primarily because we derive a much lower dust
emissivity.

Given that MM\,1 drives an energetic outflow the \tdust\ derived seems
quite realistic, while the value of $\beta$ derived is quite low. As
discussed by \citet{mookerjea2007} both theoretically and
observationally there is a large scatter in the values of $\beta$.
Based on current observations a low value of $\beta$ typically
corresponds to evolution of dust grains in circumstellar disks around
low mass protostars. Since the evolution of grains can change their
shape, size and chemical composition, it is difficult to exactly say
what causes the lower values of observed $\beta$ \citep{beckwith2000}.
\citet{miyake1993} had theoretically estimated that grainsizes of $\sim
1$~mm correspond to $\beta <1$.  Models of grain growth by collisional
coagulation predict that while grain growth could occur on very short
timescales within 10$^5$~years these grains are removed very efficiently
within a short timescale.  This however is inconsistent with observed
SEDs of classical T Tauri stars. Hence models also suggest that small
grains probably get replenished, via aggregate fragmentation through
high-speed collision \citep{dullemond2005,birnstiel2009}.  We propose
that the cause for the measured  low $\beta$-index is a partially
optically thick inner disk, which would mimic the appearance of a low
$\beta$ value.  This is plausible since many T Tauri stars have
optically thick inner disks
\citep{beckwith1990,andrews2005,andrews2007}. Hence the young disk
around MM~1, with is far more massive than a Class II disk, could well
be partially optically thick at mm-wavelengths.

We note that NGC~2023 MM\,2 has a much steeper SED than MM\,1. Our
isothermal fit gives a dust emissivity, $\beta$ = 2.08, and gives
\tdust\ = 11.4~K, suggesting that it is a cold pre-stellar core. It
has a mass of 7.4~\msun.  Although there is a faint 24~$\mu$m source,
L1630MIPS-4 (Table~\ref{tab_onlymips}) $\sim$ 5\arcsec\ from the peak
of MM\,2, there is no sign of activity in the MM\,2 core. The 24
$\mu$m source is most likely unrelated to MM\,2, although it could be
a low-mass pre-main-sequence object embedded in the MM\,1/MM\,2 cloud
core.

We have estimated the mass of the cloud harboring MM\,1 and MM\,2
using both the jiggle and scan maps observed with JCMT. For this
purpose we have smoothed the maps to a resolution of 30\arcsec\ and
then fitted a single Gaussian to estimate the total emission from the
cloud and MM\,1 and MM\,2 at 450 and 850~\micron. The flux densities
of the cloud from the scan maps and the jiggle maps were determined by
subtracting the flux densities of MM\,1 and MM\,2 in the two types of
maps (Tab.~\ref{tab_srctab}). Finally the 450 and 850~\micron\ flux
densities of only the cloud were obtained by averaging the flux
densities obtained from the scan and the jiggle maps.  From the SCUBA
observations we estimate S$_{450}$ and S$_{850}$ to be 58.2~Jy and
7.25~Jy respectively.  \citet{launhardt1996} using 1.3~mm observations
at a resolution of 30\arcsec\ obtained S$_{1300}$ to be 1.87~Jy.
Using single temperature greybody models we derive a temperature of
16.5~K and an emissivity index $\beta = 2.0$.  Using these parameters
we estimate the mass of the cloud in which MM\,1 and MM\,2 are
embedded to be $\sim$ 18~\msun.

Thus, from the present SCUBA observations we derive a total mass of
$\sim 26$~\msun\ for the entire cloud including MM~1 and MM~2.  In
contrast, \citet{johnstone2006,nutter2007}, who both identified only
one core in the MM\,1/MM\,2 cloud complex, using SCUBA studies,
derived a total mass of 14.7 and 12.7~\msun, respectively, for the
entire region.

\begin{table*}[t]
\caption{CLASS HFS fitting of selected NH$_2$D spectra. MM\,1 is at offset
(1\farcs2,1\farcs8).
\label{tab_hfs}}
\begin{center}
{\tiny
\begin{tabular}{lcccc|ccl}
\hline
\hline
Offset &  $v_{\rm cen}$ & $\Delta V$ & $\tau_{\rm m}$ & $T_{\rm R}$ & $T_{\rm ex}$ &
N$_{T_{\rm ex}=5}$ & N$_{T_{\rm ex}=10}$\\
(\arcsec,\arcsec) & km~s$^{-1}$ & km~s$^{-1}$ & & K & K &
10$^{14}$ & 10$^{14}$\\
&  & &  &  & &
cm$^{-2}$ & cm$^{-2}$\\
\hline
Disk & & & & & & & \\
(-3.7,-1.6) & 9.56$\pm$0.03 & 0.84$\pm$0.09 & 1.2$\pm$0.9 & 1.6$\pm$0.4 & 5.3 & 1.48 & 0.99\\
(-1.3, 0.1) & 9.47$\pm$0.02 & 0.85$\pm$0.05 & 1.6$\pm$0.7 & 2.0$\pm$0.4 & 5.4 & 2.02 & 1.35\\
( 1.2, 1.8) & 9.41$\pm$0.02 & 0.89$\pm$0.04 & 2.5$\pm$0.5 & 2.0$\pm$0.4 & 5.1 & 3.15 & 2.09\\
( 3.7, 3.5) & 9.34$\pm$0.02 & 0.93$\pm$0.05 & 2.7$\pm$0.6 & 2.0$\pm$0.5 & 5.0 & 3.61 & 2.39\\
( 6.1, 5.2) & 9.26$\pm$0.02 & 0.89$\pm$0.05 & 2.6$\pm$0.7 & 1.8$\pm$0.5 & 4.9 & 3.35 & 2.23\\
( 8.6, 7.0) & 9.16$\pm$0.02 & 0.80$\pm$0.06 & 2.1$\pm$0.8 & 1.7$\pm$0.5 & 4.9 & 2.39 & 1.59\\
\hline
Outflow & & & & & & & \\
( 6.4,-5.6) & 9.32$\pm$0.04 & 0.72$\pm$0.07 & 4.0$\pm$1.6 &
0.6$\pm$0.1 & 3.4 & 4.15 & 2.76\\
( 4.6,-3.1) & 9.40$\pm$0.02 & 0.78$\pm$0.07 & 3.7$\pm$0.8 &
1.1$\pm$0.4 & 4.0 & 4.11 & 2.73\\
( 2.9,-0.7) & 9.41$\pm$0.02 & 0.86$\pm$0.09 & 2.9$\pm$0.6 &
1.7$\pm$0.4 & 4.7 & 3.59 & 2.39\\
( 1.2, 1.8) & 9.41$\pm$0.02 & 0.89$\pm$0.09 & 2.5$\pm$0.5 &
2.0$\pm$0.4 & 5.1 & 3.15 & 2.09\\
(-0.5, 4.3) & 9.38$\pm$0.02 & 0.88$\pm$0.09 & 2.2$\pm$0.5 &
2.1$\pm$0.4 & 5.3 & 2.82 & 1.87\\
(-2.2, 6.7) & 9.38$\pm$0.02 & 0.84$\pm$0.09 & 1.7$\pm$0.7 &
2.3$\pm$0.5 & 5.8 & 2.02 & 1.34\\
\hline
MM~2         & 9.19$\pm$0.02 & 0.76$\pm$0.05 & 1.9$\pm$0.8 &
1.6$\pm$0.4 & 4.8 & 2.11 & 1.40\\
\hline
\hline
\end{tabular}
}
\end{center}
\end{table*}

%

\subsection{Depletion and Deuteration scenario in MM\,1 and MM\,2}

In order to derive some of the relevant physical parameters, we
analyzed the \nhtwod\ spectral profiles using the HFS method in CLASS.
This method derives the total optical depth $\tau$ and average width
of the lines in an LTE approximation along with the radiation
temperature of the multiplet, $T_{\rm R}$. Following \citet{tine2000}
we have estimated the excitation temperature $T_{\rm ex}$ for the
multiplet using the following equation

\begin{equation}
T_{\rm R} = [J_\nu(T_{\rm ex}) - J_\nu(T_{\rm bg})](1-e^{-\tau})
\end{equation}

where $J_\nu(T)$ is the radiation temperature of a blackbody at
temperature $T$ and $T_{\rm bg}$ = 2.7~K (the temperature of the
cosmic background radiation).

The total column density for \nhtwod\ is derived using the optical
depth, excitation temperature and average FWHM $\Delta v$ as follows

\begin{equation}
N = \frac{8\pi\nu_{ul}^3}{c^3} \frac{Q(T_{ex})}{g_u A_{ul}} \Delta v
\frac{e^{\frac{E_u}{kT_{\rm ex}}}}{e^{\frac{h\nu}{kT_{\rm ex}}}-1}
\tau_{\rm ul}
\end{equation}

where $\nu_{ul}$ is the frequency of the unsplit transition. The
values of the partition function, Q(T$_{\rm ex}$), used for the column
density calculations, are 4.47 and 9.37  at 5~K and 10~K respectively
\citep{roueff2005}.

Table~\ref{tab_hfs} also presents the estimated \tex\ and the column
densities calculated for \tex\ = 5~K and 10~K.  The method HFS assumes
a beam filling factor of unity, this possibly explains the rather low
excitation temperatures that we derive at all positions.  The
NGC\,2023 MM\,1 and  MM\,2 cloud core is extremely rich in deuterium,
since we observe optically thick emission at many positions
(Table~\ref{tab_hfs}). The optical depth and the column density of
\nhtwod\ along the disk shows a centrally peaked profile, with the
peak being offset from MM\,1, while along the outflow the \nhtwod\
column density increases towards the southeast. However keeping in
mind the assumptions and uncertainties in the estimation of these
column densities variation by factors of 2 to 3  are most probably not
significant. Overall, the column densities are similar in magnitude
($\sim 10^{14}$~cm$^{-2}$) to the numbers found in dark cores by
\citet{roueff2005}.

The observed high abundance of singly deuterated ammonia in MM\,1 is
suggestive of freezeout of molecules like CO, since depletion of the
heavy molecules results in an increase in the
[H$_2$D$^+$]/[H$_3^+$] and molecular D/H ratios.  Using \CeiO\ and
C$^{17}$O $J=2\to1$ data from the SCUBA archives we have investigated
the depletion scenario in MM\,1 and MM\,2. The \CeiO\ data (dashed
contours in Fig.~\ref{fig_n2023bima}) does not show any enhancement of
emission at the positions of MM\,1 and MM\,2. The C$^{17}$O emission
appears to peak at a position in between MM\,1 and MM\,2. 

We estimate the molecular hydrogen column densities, \nhtwo, for both
MM\,1 and MM\,2 using the \CeiO\ and C$^{17}$O observations. For this
we have assumed that the LTE approximation holds and have assumed
temperatures of 20~K and 11.5~K for MM\,1 and MM\,2 respectively.  We
note that the estimated column density changes only by a factor of 2
for a change in temperature from 10 to 100~K. The choice of gas
temperatures is based on the dust temperatures estimated from the
greybody fits (Sec.~\ref{sec_mm1_greyb}). Using \CeiO\ (C$^{17}$O)  we
estimate \nhtwo\ to be 1.6~10$^{22}$~\cmsq\ (2.8~10$^{22}$~\cmsq) in a
22\arcsec\ beam for MM\,1. For  MM\,2 we estimate the \nhtwo\ to be
1.7~10$^{22}$~\cmsq\ (3.0~10$^{22}$~\cmsq) from \CeiO\ (C$^{17}$O.) The
discrepancy in the \nhtwo\ calculated from \CeiO\ and C$^{17}$O column
densities is due to \CeiO\ being optically thick. The relative
abundances of \nhtwod\ at the positions of MM\,1 and MM\,2 are $\sim
10^{-9}$ (Table~\ref{tab_hfs}).

Alternatively \nhtwo\ can also be derived from the 450 and 850~\micron\
fluxes observed at MM\,1 and MM\,2. Using the observed flux densities
and dust temperatures from the greybody fits we get \nhtwo\ in the
22\arcsec\ beam to be 1.2~10$^{23}$~\cmsq\ and 2.6~10$^{23}$~\cmsq\ for
MM\,1 and MM\,2.  From dust emission within a beam of 9\farcs4,
comparable to the beam of \nhtwod, we estimate \nhtwo\ to be
2.3~10$^{23}$~\cmsq\ and 3.6~10$^{23}$~\cmsq\ for MM\,1 and MM\,2
respectively.  Thus \nhtwo\ estimated from dust continuum emission is
larger by a factor of 10 than the \nhtwo\ estimated from the CO
isotopomers. This is a clear evidence of depletion of CO and its
isotopomers by a factor of at least 8--10. This is similar to the
depletion factors observed in prestellar cores, which lie between 5.5
and 15.5 \citep{bacmann2002}.

\section{The molecular cloud south of NGC\,2023}

\begin{figure*}[t]
\begin{center}
\includegraphics[angle=0,width=16.0cm,angle=0]{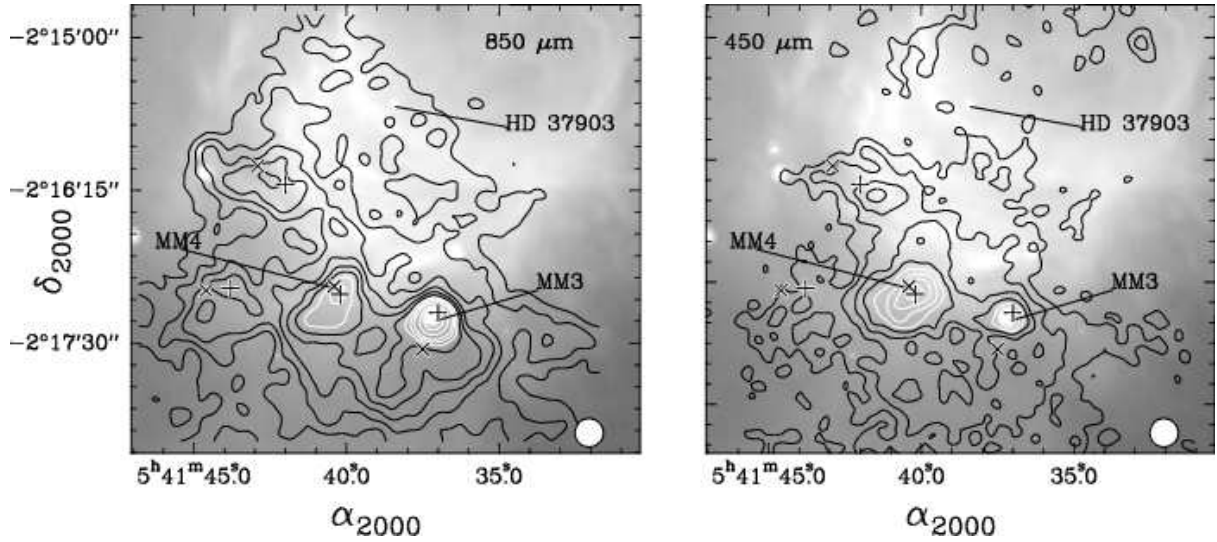}
\caption{Contours of SCUBA 850 and 450~\micron\ continuum maps of the
field surrounding NGC~2023 MM\,3 and MM~4 overlaid on the grayscale
image of Spitzer IRAC 8~\micron. The peak intensities at 850 and
450~\micron\ are 1.63 and 6.47~Jy/beam respectively.  The contour for
the 850~\micron\ map are at 0.2, 0.3, 0.4, 0.5, 0.6, 0.7, 0.8, 1, 1.2,
1.4~Jy/beam.  The contours for the 450~\micron\ map are at  1.2, 1.75,
2.5, 3.25, 4, 4.75, 5.5, and 6.25~Jy/beam.  The beamsizes at 450 and
850~\micron\ are shown at the right bottom corner of each panel. ``+"
and ``x" signs mark the sub-mm sources detected in this region by
\citet{johnstone2006} and \citet{nutter2007}, respectively.
\label{fig_n2023MM3}}
\end{center}
\end{figure*}

The reflection nebula NGC\,2023 is bounded to the south and southeast
by a dense, extended molecular cloud coinciding with the ridge of
strong PAH emission seen in the IRAC and MIPS images.  At 850 $\mu$m
the cloud has a size of $\sim$ 98\arcsec\ $\times$ 78\arcsec\ PA
$\sim$ 34\degr\ measured at the 3-$\sigma$ contour level with an
integrated flux  of  $\sim$ 16.7 Jy. Some faint, diffuse dust emission
is also seen toward the reflection nebula. The strongest dust emission
is just outside or overlapping with the bright PDR ridge seen in IRAC
and MIPS images and the ionized carbon (C$^+$) ridge. The latter was
mapped by \citet{wyrowski2000}  with the VLA in the C91$\alpha$
recombination line.  At high resolution the dust emission breaks up
into several cores (Figure~\ref{fig_n2023MM3}). Both
\citet{johnstone2006} and \citet{nutter2007}   identified four cores
in this cloud. We can securely identify only two at both 450 and
850~\micron. One of them (MM\,3) is compact and coincides with a
bright mid-IR source, which has the colors of a Class I/0 source
(Section~\ref{sec_YSO_population}).  The other core, MM\,4, is much
more extended and has no near- or mid-IR counterpart. Even though we
see 850 $\mu$m emission at the approximate positions of the other two
cores seen by \citet{johnstone2006} and \citet{nutter2007}, they have
no counterparts at 450 $\mu$m, and we do not consider them as cores.
Since the \citet{johnstone2006} map goes deeper, our failure to see
these cores could be due to our limited sensitivity. It is, however,
possible that what \citeauthor{johnstone2006} and
\citeauthor{nutter2007} identify as cores, are structures that break
up into smaller clumps at the higher angular resolution at
450~\micron, especially since neither core can be identified in any
molecular line map.

The molecular gas in the PDR region is hot. The $^{13}$CO $J = 2 \to
1$ map shown in the \citet{johnstone2006} paper and re-reduced by us
from the JCMT archive show  $^{13}$CO brightness temperatures of 50~K
towards MM\,4, which means that the gas temperature is $\geq$ 50 K.
Such high gas temperature agrees well with the color temperatures of
40--50~K derived from far infrared observations at 50 and 100 $\mu$m
\citep{harvey1980} one arcminute south and west of HD~37903. The color
temperatures determined between 160 and 100 $\mu$m are somewhat lower,
30--40 K. The latter color temperatures agree well with
\citet{mookerjea2000}, who determined dust temperatures of $\sim$30 K
from 205 and 138 $\mu$m over most of the cloud south of NGC\,2023. The
hot gas is probably restricted to a narrow zone between the reflection
nebula and the molecular cloud. If we assume a temperature of 30 K
for the cloud as a whole, we find a cloud mass of 30 \msun\ from
C$^{18}$O J = $2 \to 1$, while the mass seen in 850 $\mu$m dust
emission, which only probes the densest parts of the cloud core, is
only $\sim$ 14.3 \Msun.  Both mass estimates are for the
continuum and line fluxes from a region extending over 98\arcsec$
\times$ 78\arcsec. The total continuum emission over this region is
16.7~Jy, from which 1.3 and 1.6~Jy due to the emission from MM\,3 and
MM\,4 are subtracted to get a flux of 13.8~Jy for the cloud only. The
average dust temperature is assumed to be 30~K, i.e. similar to what
we assumed for  C$^{18}$O.  In this case the cloud mass derived
from C$^{18}$O is about twice as large than what we derive from dust
emission. If C$^{18}$O is undepleted, we would expect to derive a
larger mass from the C$^{18}$O  emission than from  dust emission,
since  C$^{18}$O is also excited in low density gas, while the dust
emission only traces the very dense gas in the cloud core. If the gas
temperature is even higher, which could well be the case, it would
make the difference in mass even larger. This strongly suggests that
in contrast to the MM\,1/MM\,2 region, \CeiO\ is not depleted in the
MM\,3/MM\,4 region of NGC\,2023.

\begin{figure*}[htbp]
  \centering
  \begin{minipage}[b]{5.5 cm}
    \includegraphics[width=5.5cm]{bm_fig9a.eps}  
  \end{minipage}
 \hspace{0.5cm} 
  \begin{minipage}[b]{5.5 cm}
    \includegraphics[width=5.5cm]{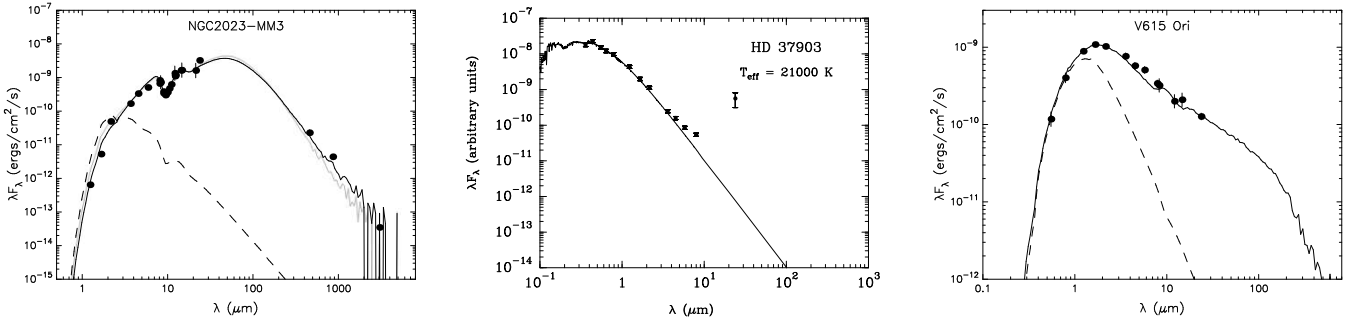}  
  \end{minipage}
 \hspace{0.5cm} 
  \begin{minipage}[b]{5.5 cm}
    \includegraphics[width=5.5cm]{bm_fig9c.eps}  
  \end{minipage}
\caption{SED fits for NGC~2023 MM\,3, HD\,37903 and V615~Ori. The
filled circles indicate the measured fluxes and uncertainties. For
MM~3, the thin black line represents the best-fitting SED, and the
gray lines represent the all other acceptable ($\chi^2$--$\chi^2_{\rm
best} <$ 10) YSO fits. For V615~Ori the thin black line shows the best
fitting model which reproduces the spectral class of the source.
The dashed line shows the stellar photosphere corresponding to
the central source of the best fitting model, as it would look in the
absence of circumstellar dust (but including interstellar extinction).
For HD~37903 the model for a stellar photosphere corresponding to a
B2V star is shown to conclusively demonstrate the infrared excess
beyond 4~\micron.}
\label{fig_sedfits}
\end{figure*}

\section{The NGC2023 MM\,3/MM\,4 region
\label{sec_mm3_mm4}}

\subsection{MM\,3}

MM\,3, coincides with an extremely red IR source, originally
discovered by \citet{depoy1990}, \#3 in their list of near-IR
sources associated with NGC\,2023.  The source is also in the 2MASS
(although not listed by SIMBAD) and the MSX point source catalogues.
We detected it in all IRAC bands and in the MIPS 24~\micron\ image.
\citet{wyrowski2000} detected MM\,3 at 3~mm with a flux density of
36$\pm$7 mJy, although they interpreted the dust emission to originate
in an extended cloud core surrounding the near-IR source detected by
\citet{depoy1990}.  We find the dust emission to be more compact;
at 450 $\mu$m the emission appears unresolved. The 850 $\mu$m size
quoted in Table~\ref{tab_srctab} is therefore an upper limit. The
excellent spatial agreement between our sub-millimeter position for MM\,3 and
the 2MASS and IRAC position ($\leq$ 0\farcs5) therefore supports the
classification from the IRAC colors, which suggest that MM\,3 is a
Class I/0 protostar. 

Even though MM~3 is not associated with free-free emission
\citep{reipurth2004}, it may excite the HH objects HH\,5,  HH\,4, and
HH\,1~A--C, which were discovered by \citet{malin1987}.  Examination
of the $^{13}$CO $J = 2 \to 1$ map shows faint high-velocity wings
near the position of MM\,3 and faint red-shifted wings to the east of
the source. If the outflow is bipolar the blue-shifted molecular
outflow would therefore be to the west, i.e. in the direction towards
the HH objects. There is, however, considerable velocity structure in
$^{13}$CO in the NGC\,2023 PDR region, and this is by no means a clear
identification. L1630MIR-62  (Sellgren's star C, see
Section~\ref{sec_onlymir}) is somewhat better aligned with the HH
objects. Another possibility is a very red near IR source, L1630NIR-15 ,
which is within 1\arcsec\ of HH\,1\,A.

Because MM\,3 lies within the boundary of the reflection nebula, which
is very bright in the far infrared,  there are no flux density
estimates or even stringent upper limits at 100 $\mu$m, which could
help us constrain the temperature of the dust emission. However,
MM~3 is a visible IR object, presumably warmer than 20 K (MM~1) but
probably not too hot. We therefore assumed  a dust temperature of
25~K and estimated the dust emissivity and dust mass with a single
temperature greybody fit using the observed sub-millimeter and
millimeter flux densities. We obtain a  dust emissivity index of 0.8
and a dust mass of 0.33~\Msun. Although the dust emissivity is on the
low side, most protostellar objects show dust emissivities in the
range of 0.7--1.5, and often similar ``disk'' masses as well
\citep{dent1998,froebrich2005}, re-inforcing our hypothesis that the
dust emission originates in a disk/envelope around a young protostar.

Below we analyze in more detail what we can learn about MM\,3 with the
assumption that the dust emission originates in a disk or compact
envelope surrounding the protostar.

\subsection{Nature of MM\,3 from SED fitting}

For a more detailed model which explains the emission from MM\,3 over
a wide range of wavelengths we have combined the SCUBA data with the
2MASS, {\em Spitzer} (IRAC, MIPS, and IRS) and MSX data and
measurements at 3~mm (BIMA). 
\citet{robitaille2007} have created an archive of two-dimensional (2D)
axisymmetric radiative transfer models of protostars calculated for a
large range of protostellar masses, accretion rates, disk masses and
disk orientations.  This archive also provides a linear regression
tool which can select all model SEDs that fit the observed SED better
than a specified $\chi^2$.  Each SED is characterized by a set of
model parameters, such as stellar mass, temperature, and age, envelope
accretion rate, disk mass, and envelope inner radius.  We have used
this online tool to generate models which fit the observed SED for
NGC~2023\,MM\,3. The models assume that stars of all masses form via
free-fall rotational collapse to a disk and accretion through the disk
to the star; thus the envelope accretion rate sets the envelope mass.
Since the distance to the source is somewhat uncertain, we have
explored a range of distance between 0.3 and 0.5~kpc and \av\ between
0 and 100~mag. Aperture radii used for the fits are 30\arcsec\ for the
MSX data, 5\arcsec\ for the IRAC and IRS data, 9\arcsec\ for the MIPS
24~\micron\ data, 14\arcsec\ for the
SCUBA data and 13\arcsec\ for the 3~mm BIMA data.  Left panel of
Fig.~\ref{fig_sedfits} shows the SED of the best-fitting model for
MM\,3 and one more model which fits the data with
$\chi^2$--$\chi^2_{\rm best} < 10$.

We have obtained the average values and variances of the fitted
parameters by taking a weighted average of the two models for which
$\chi^2$--$\chi^2_{\rm best} < 10$, in a manner similar to
\citet{simon2007}. Both these models correspond to a distance $D$ =
350~pc.  Here we present the parameters obtained from the two
best-fitting models, viz., the stellar mass ($M_\ast$), temperature
($T_\ast$) and luminosity ($L_\ast$), the envelope mass $M_{\rm env}$,
disk mass $M_{\rm disk}$,  and the envelope accretion rate
$\dot{M}_{\rm env}$. The numerical values are :
$M_\ast=0.88\pm0.22$~\msun, $L_\ast$=21$\pm1.4$~\lsun,
$T_\ast$=3930$\pm$168~K, $M_{\rm env}$ = 0.24$\pm0.02$~\msun, $M_{\rm
disk}$=0.007$\pm$0.004~\msun, and $\dot{M}_{\rm env}$ =
3.3~$10^{-5}\pm8.0~10^{-6}$~\msun\ yr$^{-1}$.  The envelope mass
matches reasonably well with the mass estimated from the simple
greybody fit.  The SED fit is very well constrained in the MIR
wavelengths owing to the IRS datapoints. As a result we are able to
arrive at only two models (as opposed to a family of models), which
reproduce most of the observed features reasonably well, with the best
fit model fitting the longer wavelength data significantly better.
We note that there is a substantial discrepancy in the flux
densities measured in the MSX E-band and MIPS 24~\micron, although
both bands have very similar wavelengths. The best-fit model presented
here underestimates the MIPS 24~\micron\ flux. We have checked that
the quality of the fit does not change significantly by not
considering the MSX fluxes, since in the MIR the fit is primarily
constrained by the IRS fluxes. Thus, only more datapoints in the
far-infrared region will enable us to arrive at a better constrained
fit particularly at the longer wavelengths of MIR to FIR.  


\subsection{MM\,4}
\label{results_mm4}

Based on \thCO J= 2$\rightarrow$1 brightness temperatures $\geq$ 50~K,
MM\,4 appears warmer than MM\,3. However in the absence of
far-infrared fluxes it is not possible to derive the dust temperature
by performing an isothermal fit to the 450 and 850~\micron\ fluxes
unless the value of $\beta$ is accurately known. We therefore
considered two isothermal fits constraining \tdust\ to 30~K and then
to 50~K. In both cases we obtained a fitted value of 1.5 for $\beta$.
The masses corresponding to temperatures of 30 and 50~K are
0.87~\msun\ and 0.46~\msun\ respectively.  Irrespective of the
temperature, the value of $\beta$ for MM\,4 is lower than  $\beta =
2$, which is  typically seen in colder pre-stellar cores. This could
be due to heating of the core from the outside, which would suggest
that the core is not well described by a single temperature. Another
possibility is that the dust has undergone processing and is different
in the PDR region.  Which explanation is the right one, or whether it
is a combination of both, will have to wait until we can observe MM\,4
with higher spatial resolution at several wavelengths in
sub-millimeter and far infrared.

\section{NGC\,2023~MM\,5 region}
\begin{figure*}[ht]
\begin{center}
\includegraphics[angle=0,width=16.0cm,angle=0]{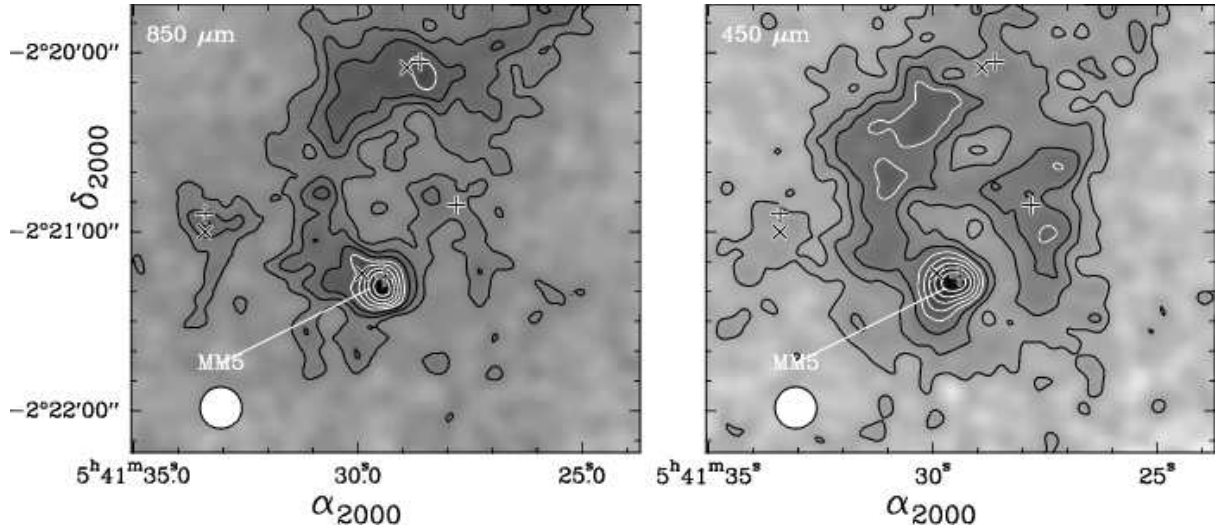}
\caption{Gray-scale 850 and 450~\micron\ continuum maps overlaid with
contours of the field surrounding NGC~2023 MM~5.  The peak intensities at
850 and 450~\micron\ are 0.88 and 5.38~Jy/beam respectively. The
contours for both the 850 and 450~\micron\ maps are at  20 to 100\% of
the respective peak intensities in steps of 10\%.  The beamsizes at 450
and 850~\micron\ are shown at the right bottom corner of each panel.
``+"  and ``x" signs mark the sub-mm sources detected
in this region by \citet{johnstone2006} and \citet{nutter2007},
respectively.
\label{fig_n2023mm5}}
\end{center}
\end{figure*}

MM\,5 is a compact object in the southern part of an arc or a
shell-like structure in the southern part of the dust ridge
(Fig.~\ref{fig_n2023mm5}).  There are no optical, infrared or radio
sources associated with MM\,5 region. Examination of our $^{12}$CO and
$^{13}$CO  $J = 2 \to 1$ maps show no distinctive features in this
area and the $^{12}$CO brightness temperature is $\sim$ 15 K,
suggesting that the dust emission is associated with a cold prestellar
core.  An isothermal greybody fit to the integrated flux densities
suggest a dust temperature of  12.2 K, $\beta$ = 2.2, and a mass of
4.3 \Msun.  This core was also picked up by \citet{johnstone2006} and
\citet{nutter2007}, who find approximately the same mass.  Both
identified the dust emission in the northern part of the arc as a
separate cloud core (Fig.~\ref{fig_n2023mm5}).  \citet{johnstone2006}
identified an additional three sub-millimeter ``clumps'' in this
region. It is impossible to judge whether all of these are true cores,
without observing the region in a molecular line tracer sensitive to
dense, cold gas.

\section{Other sources of interest in the NGC 2023 region
\label{sec_onlymir}}

Comparison of high resolution mid-infrared and sub-millimeter data
with medium resolution far-infrared data suggests that the emission
from L1630, the molecular cloud associated with NGC~2023 is primarily
extended and diffuse emission. However we also find a number of
embedded Class I and II objects (see Sec.~\ref{sec_YSO_population}),
some of which we would have expected to detect in the sub-millimeter,
since they have clear IR excess and at least some are expected to be
surrounded by a circumstellar disk and/or envelope.  Yet we detect
only one MIR source in the sub-millimeter. Below we highlight some of
the stars for which our IRAC and MIPS photometry provide new
information about their nature, {\em viz.,}  HD~37903, V615~Ori,
Sellgren's star C, and NGC\,2023-VLA3. 

{\bf HD\,37903 or L1630MIR-66}  illuminates the reflection nebula NGC\,2023
located in the northern part of  the L\,1630 molecular cloud.  SIMBAD
lists the spectral type as B1.5 V. This star has also been classified
as B2 V  \citep{stebbins1940,crawford1958}.
\citet{warren1977} assign it a spectral class between B 1.5V to B2 V.
and note that it may be variable in V, which is confirmed by examining
the measurements in the the SIMBAD data base. \citet{stephenson1986}
identified HD\,37903 as a weak H$\alpha$ emission line star in an
objective prism survey of the northern sky. Yet \citet{hernandez2005}
did not classify it as a  Herbig Ae/Be star in their study of early
type stars in nearby OB associations. Since HD\,37903 illuminates a
reflection nebula, is an early B-star, is photometrically variable and
as we show below, has an infrared excess, it appears to have all the
characteristics of a Herbig Ae/Be star \citep{the1994}. Although it is
a very bright star (V = 7.8$^m$) and extremely well studied, almost
all spectral classifications of the star are based on photometry.
There are, however, several long integration UV spectra of excellent
quality available in the archives of the International Ultraviolet
Explorer (IUE), which enabled us to perform an improved spectral
classification. We therefore looked at the short wavelength high
dispersion (SWP) spectra in the IUE archive and used the spectral
classification of B stars  developed by \citet{rountree1991} for
ultraviolet spectra to derive a more accurate spectral classification.
This comparison shows that the star is definitely later than B1.5 and
earlier than B2.5, i.e. a B2 V star is a very good match.  We
therefore conclude that HD\,37903 is a Herbig Ae/Be star of spectral
class B2 Ve.

To verify that the star has an infrared excess we compared the
observed infrared fluxes to the Kurucz stellar atmosphere model
corresponding to solar metalicity, log[g]=3.0 and an effective
temperature of 21000~K (Fig.~\ref{fig_sedfits} {\em middle}). We find
that while the optical and NIR fluxes are satisfactorily reproduced by
the Kurucz model, the IRAC fluxes from 4.5~\micron\ show clear
infrared excess compared to the model and a very large excess at
24~\micron\ possibly indicating the presence of a protostellar disk.
Although the star is clearly seen at 24~\micron, the measured flux
density is very uncertain due to the intense PDR ridge just south of
it. Since HD\,37903 illuminates a faint \HII\ region
\citep{wyrowski2000}, some of the 24 \micron\ emission also originate
in the \HII\ region rather than from a disk.  Although HD\,37903 has
an IR excess, we see no trace of it in the sub-millimeter continuum.
This is not surprising, since disks around early B-stars are likely to
be short lived. We do not know of any early B-star, that illuminates a
large reflection nebula and still would be associated with a cold
protostellar disk \citep[see e.g.][]{fuente1998,fuente2001}.

{\bf V615~Ori or L1630MIR-52} is a heavily reddened T Tauri  star with
spectral type of  G7 III--IV \citep{malin1987}. It is one of the few
PMS stars in the field that  is a genuine IRAS source (IRAS
05388$-$0224) and not a false detection\footnote{IRAS 05384$-$0229 and
IRAS 05386$-$0229 in the Horsehead nebula; IRAS 05385$-$0222, IRAS
05384$-$0219, and IRAS 05385$-$0212 in the IC434 ionization front;
IRAS 05392$-$0214, IRAS 05390$-$0214, and IRAS  05391$-$0217 (near
HD\,37903) in NGC2023 are all spurious detections, since they are not
seen in any IRAC bands nor in the MIPS 24 $\mu$m image}, with flux
densities of 1.45 Jy
at 12 $\mu$m and 1.67 Jy at 25~\micron. It is listed in the  2MASS
point source catalogue and we detect it in all IRAC bands and in the
MIPS 24~\micron\ image. 
\citet{malin1987}
quote an observed B-V color of 2 (suggesting E$_{B-V}$ = 1.2).  In the
General Catalogue of Variable stars \citep{kukarkin1971} the listed V
magnitudes 
vary between 13.8 and 15.9.  These values  agree well with more recent
data of \citet{droege2007}, who give $V=$13.11$\pm$0.26 and
$I=$10.99$\pm$0.16. The star is also an X-ray source, detected by the
XMM-Newton telescope (XMM-Newton 2nd Incremental Source Catalogue).  Using the
optical and infrared flux densities and the online SED fitter we
find 7 models for which $\chi^2$--$\chi^2_{\rm best} < 0.6$.
However 6 out of the 7 models have stellar
temperatures $>6000$~K, which is inconsistent with the known spectral
type of V615~Ori. We thus present here the only model which
corresponds to a stellar temperature agreeing with previous
observations. The parameters for this model are: Distance = 417~pc,
$M_\ast$ = 4.0~\msun, $T_\ast$ = 5546~K, $L_\ast$ = 62.7~\lsun,
$M_{\rm env}$ = 0.056~\msun\ and $M_{\rm disk}$ = 0.006~\msun. If we
use the power-law dependence between disk mass and 850~$\mu$m flux
density derived from 153 T Tauri stars in Taurus-Auriga
\citep{andrews2005}, we find that the modeled disk and envelope mass
would result in an 850 $\mu$m flux density of $\sim$ 4 mJy, which is
below our detection limit. It is therefore not surprising that we did
not detect the star in our SCUBA observations nor would it have been
detected in the deeper 850 $\mu$m image by \citet{johnstone2006}. 

\citet{malin1987} found two HH objects (NGC\,2023-HH\,2, and HH\,3)
southeast of V615~Ori. The proper motions of HH\,2, however, suggest
that the exciting source is south or southeast of HH\,2 rather than in
the direction of V615~Ori. There is a very red MIR source, L1630MIR-51
(Table~\ref{tab_iracsrc}), south-east of HH\,2, which could be the
exciting star for HH\,2, and possibly HH\,3. L1630MIR-51 has very strong
infrared excess and colors of a Class II/I object, which indicate that
it is very young and deeply embedded.  Examination of the large CO $J
= 2 \to 1$ map to which we have access, indicates that V615~Ori may
drive a weak, extended CO outflow. A position-velocity plot through
the star and HH\,3 shows faint blue-shifted emission to the
south-east, i.e. in the direction of HH\,3, and red-shifted emission
to the northeast, with a reversal of the line wings approximately at
the position of V615~Ori. Nothing stands out in the CO map at the
position of HH\,2 or L1630MIR-51.  We therefore believe that HH\,3 is
excited by V615~Ori, although we caution that the identification is
tentative and has to be confirmed with deeper observations with better
spatial resolution. HH\,2 may be part of the V615~Ori outflow,
although we cannot exclude that this HH complex is excited by L1630MIR-56,
which is $\sim$ 30\arcsec\--40\arcsec\ south east of HH\,2~A,B.

{\bf Sellgren's star C (L1630MIR-62, HBC\,500}\footnote{Both Sellgren's
star C \citep{sellgren1983} and HBC\,500 have incorrect coordinate
entries in the SIMBAD data base; neither corresponds to a visible star
or IR source.}) was also found to be a T Tauri star by
\citet{malin1987}, but with much stronger H$\alpha$ emission than
V615~Ori. However, since the star is heavily obscured, they were able
to only conclude that the star has a spectral type of late G or early
K and a luminosity class V, based on  weak Na I and Mg I absorption.
\citet{weaver2004} also found it on a very deep CTIO objective prism
plate from 1976. In their catalogue it is number 54, and they
characterize the star as having medium strength H$\alpha$ emission
with weak continuum. The star is rather bright in IRAC and MIPS
images, L1630MIR-62, (Table~\ref{tab_iracsrc}), and has MIR  colors
of a Class II object.


{\bf NGC\,2023-VLA\,3 or L1630MIR-46} was the strongest radio source
($\sim$ 50 mJy) in the 3.6 cm VLA survey by  \citet{reipurth2004}. It
has no optical or sub-millimeter counterpart, but is detected in all IRAC
bands and also in the 24 $\mu$m MIPS image (Table~\ref{tab_iracsrc})
with colors of a Class II or borderline Class I source. The flux
density is too high to be free-free emission from an ionized jet. It
is almost certainly gyrosynchrotron emission, which is most often found
in weak-line (naked) T Tauri stars, but sometimes in young stellar
objects with strong magnetic fields \citep{andre1992}. Since the star
is an X-ray source (XMM-Newton 2nd Incremental Source Catalogue), it
is likely to be have magnetic activity.

{\bf Additional sources:} \citet{weaver2004} identified seven additional
H$\alpha$ emission line stars within our field of view, WB\,53 -- 59.
WB\,53 is most likely the Herbig-Haro object HH\,5 all the others have
MIR counterparts and all but two, WB\,57 and WB\,59, have clear
IR-excess.  VLA\,2 has no near- or mid-IR counterpart and is most
likely an extragalactic source.

\section{Discussion}
\label{sec_discussion}

\subsection{The YSO population
\label{sec_ysopop}}

Extracting photometry of ``faint'' stars in a bright knotty reflection
nebula like NGC\,2023 is rather involved. The {\it Spitzer} photometry
package APEX recovered far more sources than what we present in
Table~\ref{tab_iracsrc}. However, many of the ``detections'' had blue
[5.8]--[8.0] colors and very red [4.5]--[5.8] colors, i.e. they
appeared non-stellar. Visual inspection of each source showed that
they all appeared nebulous and fuzzy. It is quite likely that some of
them may contain real stars. The PSF photometry, however,  was
definitely dominated by nebular emission and/or absorption present in
the broad IRAC filters, and we therefore rejected them. The final list
may still contain some spurious sources, but if there are, they are
very few.

With a total of 73 YSOs identified in the NGC~2023/L~1630 the sample
is sufficiently large to visualize the distribution of the YSOs
(Fig.~\ref{fig_n2023l_850}).  We find that the majority of them, $\sim
23$ stars, are located in the southern part of NGC\,2023. There is
another 7--9 young stars in the sub-millimeter ridge west of
NGC\,2023, in which the Class 0 protostar, MM\,1, is embedded. If we
consider these to belong to the NGC\,2023 cluster, we therefore have a
total  of 32 YSOs. We also find a string of YSOs (all Class II) along
the whole submillimeter ridge, which runs parallel to the bright PDR
rim IC\,434.  A slightly larger fraction of the YSOs  lie on the
western side of this dust ridge, i.e.  the side facing the large
IC\,434 \HII\ region.  The rest of the YSOs lie in the bright rim
created by IC\,434 or slightly inside  IC\,434.  Here we observe two
distinct groups. One group is associated with the Horsehead nebula,
where we see 7 sources just outside the head region and another two
down the neck, where the Horsehead nebula merges into the L\,1630
molecular cloud.  Another small group is seen in the northwestern part
of our field. The latter group is centered around an eroded elephant
trunk structure, which is now much smaller than the Horsehead.  A few
PMS stars (including VLA\,3) are scattered all over the L\,1630 dark
cloud, suggesting that  occasional star formation can occur in any
dense cold cloud core without any external triggering.  Of the 5 bona
fide Class I/0 stars from Table~\ref{tab_iracsrc} and the 10 very
young MIPS-only sources,  a total of 9 are concentrated  in the
southern part of NGC\,2023 and near or in the MM\,1 cloud core. The
remaining 6 are distributed in a scattered fashion close to the
Horsehead nebula and the T Tauri star V615~Ori.  The strong
concentration of Class I/0 sources near NGC\,2023 suggests that this
is where the youngest stars are and where star formation may still be
on-going.

Our YSO sample  includes all  the H$\alpha$ emission line objects from
the \citet{weaver2004} survey. However the two emission line objects,
WB\,57 and WB\,59, which are inside the NGC\,2023 reflection nebula
and are to faint to be detected at 8.0~\micron, but are readily
detected at 3.6 and 4.5~\micron. Since both lie in areas of intense
PAH emission, it is possible that the neutral or slightly blue [3.6] -
[4.5] color is due to intense PAH emission in the 3.6~\micron\ band.
As we have no way of confirming this at the present, we have counted
them as members of our candidate YSO population.

Our study re-enforces the view that a large \HII\ region like IC\,434
expanding into dense surrounding cloud complexes triggers the formation
of new stars. The densest regions in the surrounding molecular clouds,
like the Horsehead nebula, can even form small clusters. The Horsehead
may even form additional stars, because we found one Class I/0 source
near the base of the Horsehead. There is also a bright, extended
pre-stellar core in the neck of the Horsehead, which is seen as an
infrared dark cloud in IRAC and MIPS images. The pre-shock from the
\HII\ region can compress molecular gas even further into the cloud and
trigger another wave of star formation. This is demonstrated by all the
young stars we find in submillimeter ridge, which runs parallel to the
IC\,434 boundary, but almost  0.8 pc  from the edge of IC\,434.

A comparison of our results with the recent study to derive an
infrared census of YSOs within a 7\arcmin$\times$7\arcmin\ region
around the Horsehead nebula \citep{bowler2009}, shows a very good
agreement with the results we get from IRAC photometry. We have
identified almost all the sources with an IR excess, which
\citet{bowler2009} present in their Table~1.  The sources which we do
not detect are B33-13, B33-20,B33-27, B33-29, and B33-30 all of which
were selected on the basis of their NIR colors, a criterion which does
not match with our MIR-based approach. The classification of these
sources derived by \citet{bowler2009} also match with the classes
derived in this paper. Although \citet{bowler2009} report a
significant 8~\micron\ excess for B33-31 and classify it as a
bona-fide YSO, we do not detect the source at all at 8.0 ~\micron.  We
do retain it as a candidate YSO. Since we have also analyzed MIPS data
for the region, we do identify a larger number of candidate YSOs than
seen by \citet{bowler2009}.

The present {\em Spitzer}-based study has enhanced the number of YSOs
in the young star cluster associated with NGC\,2023, which has
HD\,37903 as its most massive member.  \citet{depoy1990} found 16
young stars in this cluster based on deep near-IR imaging. We find
that two of them are heavily reddened foreground stars, and the true
number is therefore 14. In this study we find $\sim$ 32 or possibly as
few as 22 YSOs in the star cluster associated with NGC~2023.  Even
with  our larger estimate of the number of YSOs in the NGC\,2023
cluster, we are not even close to the number of stars, 50,
\citet{depoy1990} had estimated to be surrounding HD\,37903.  Their
estimate was based on the assumption that even a small cluster follows
a Salpeter or Miller \& Scalo IMF,  interstellar mass function
\citep{miller1979}.  Several investigations, especially from groups
that study clusters in star forming galaxies, show that there can be
substantial stochastic fluctuations when one deals with small  number
statistics \citep[see e.g.,][]{cervino2003,carigi2008}. However, we do
know that we have missed PMS stars at the faint end, i.e. very low
mass stars. If we add stars, which were not seen by 2MASS, but which
are very red ($\geq$ 0.3) based on their [3.6]--[4.5] colors, we find
another 8 stars south of the reflection nebula, one to the north west
of HD\,37903 inside the reflection nebula, and another 7 in the
sub-millimeter ridge. Most of these are likely to be low mass PMS
stars.  The true, but still conservative number for the NGC\,2023
cluster is therefore $\sim$ 32 stars.  If we account for stochastic
fluctuations and also take into consideration that this a very young
cluster, where star formation is still going on, there is nothing
anomalous with the mass distribution of these young stars.  Some  of
the low-mass stars may still be in the process of forming or be deeply
embedded in the molecular cloud core.  They may therefore not  even be
detectable by deep mid-IR observations, like the IRAC observations
that we have employed in this study.

\subsection{The sub-millimeter and protostellar candidates
\label{sec_disc_submm}}

SCUBA maps at 450 and 850~\micron\ reliably detect only 5
sub-millimeter sources, MM\,1--5. Of these sub-millimeter cores, MM\,1
has been shown to be a Class 0 object by \citet{sandell1999}. Based on
isothermal grey-body modelling we find that MM\,1 has a rather low
value of $\beta$. Since MM\,1 is known to drive an energetic outflow,
it must have an actively accreting disk. It therefore seems that the
inferred low $\beta$ value is unlikely to be due to the formation of
large grains. We argue that the low $\beta$ arises due to the inner
disk being partially optically thick.  We have imaged the protostellar
disk associated with MM\,1 using interferometric millimeter
observations of the \nhtwod\ emission. However based on the observed
velocity structure we conclude that the \nhtwod\ emission appears to
trace rotation in the surrounding envelope rather than the
protostellar disk. The MM~1/MM~2 region coincides with an IR dark
cloud and the presence of two faint MIPS sources  suggests that stars
have already formed in the cloud core. MM\,2 and the surrounding cloud
core is extremely dense and cold and is not evident in single dish
C$^{18}$O and C$^{17}$O, while NH$_2$D shows a very similar morphology
to that of the dust continuum, especially for the cold, dense, dark
cloud core surrounding MM\,1 and MM\,2. Analysis of column densities
derived from C$^{18}$O and C$^{17}$O J = $2 \to 1$ compared to column
densities estimated from the dust emission suggest that CO is strongly
depleted in the cold core, presumably because a large fraction of the
CO is frozen onto dust grains.  \nhtwod, on the other hand, appears to
be strongly enhanced in the cold cloud core. The mass of the gas of
MM\,2 and the cloud core surrounding MM\,1 and MM\,2,  $\sim$ 25
\Msun, is certainly sufficient to form additional low-mass stars in
this cloud core.

The sub-millimeter sources MM\,3 and MM\,4  lie next to the NGC~2023
reflection nebula, which is where we see the highest concentration of
young stars. Here the surrounding molecular cloud is much warmer and
we detect no CO depletion. MM\,3 is a compact sub-millimeter source,
which coincides with a bright mid-infrared source, L1630MIR-63, which has a
very strong infrared excess and which we classify as an extreme Class
I source based on color-color diagrams. In the millimeter and
sub-millimeter is looks almost identical to MM\,1, although it is not
clear whether it drives an outflow and it does not have any observable
free-free emission. The  dust emissivity index, $\beta$ = 0.8, deduced
for MM~3 is almost identical to MM\,1 and although the disk/envelope
mass is only about a third of the disk around MM\,1, both stars have
very massive disk. Yet MM\,3 is detected even in the near-IR, while
MM\,1 is not even seen at 24 $\mu$m, suggesting that MM\,3 is a more
evolved star than MM\,1. It is, however, possible that the main reason
why we do not detect MM\,1 in the mid-IR is because we see the disk
nearly edge on. If the disk surrounding MM\,3 is more edge on, and
therefore not blocking the light from the central protostar, both
could be in a very similar stage of their evolution. MM\,4 is the most
extended dust core we identified in our sample. It differs from
typical pre-stellar cores, because it is relatively warm and it also
has a slightly anomalous $\beta$-index.  MM~5 is a cold prestellar
core with both temperature and $\beta$ characteristic of most
pre-stellar cores.  

As discussed briefly in Secs.~\ref{results_mm1_mm2} and
~\ref{sec_mm3_mm4} we find substantially different masses for almost
all of the sub-millimeter cores that overlap with the sample of
\citet{johnstone2006} and \citet{nutter2007}, except MM\,5, (SMM
J054149-02212 \citep{johnstone2006}, and OrionBS-541299-22114
\citep{nutter2007}).  For this sub-millimeter core all three studies
find about the same mass, $\sim$ 3--4 \Msun. Here we review why there
are such large difference in masses estimated for the cores MM\,1-- 4.

\citet{johnstone2006} and \citet{nutter2007} estimate very similar
masses for the MM\,1/MM\,2 region, $\sim$ 15 \Msun\
\citep[SMM~J054141-02180][]{johnstone2006} and $\sim$ 13 \Msun\
\citep[OrionBS-541257-21809][]{nutter2007}, while we show that this
``core'' breaks up into two sub-millimeter cores embedded in a very
dense, cold cloud. MM\,1 coincides with a Class 0 protostar, while
MM\,2 appears to be a pre-stellar core
(Section~\ref{results_mm1_mm2}). They differ in temperature and have
very different SEDs. MM\,2 has a $\beta \sim$ 2, while MM\,1 has a
very flat SED with $\beta$ = 0.7.  If we add the masses for MM\,1 and
MM\,2 we get $\sim$ 8 \Msun, only about half of the mass found by
treating MM\,1 and MM\,2 as a single core. For MM\,3, which we also
identify as a Class 0/I protostar, with an SED similar to MM\,1, the
differences in mass are large. We find 0.3 \Msun, while
\citet{johnstone2006} determine a mass for  of $\sim$ 16 \Msun\ (core
SMM~J054162-02172), and \citet{nutter2007} find 8 \Msun\ (core
OrionBS-541375-21733). Here the large difference is in part due to
MM\,3 having a very flat SED, but also because both groups deduce a
much larger size for MM\,3 than we do. We find it point-like and
unresolved (Table~\ref{tab_srctab}). For MM\,4 \citet{johnstone2006}
(SMM~J054167-02171) and \citet{nutter2007} (OrionBS-541404-21702)
obtained masses of 8 and 11.0\Msun, respectively,  which is about 10
times higher than what we find. In this case the discrepancy primarily
arises due to the differences in the derived source size and assuming
normal  dust properties, while we find that the dust emissivity
differs from that of a typical pre-stellar core
(Section~\ref{results_mm4}).

We attribute the primary source of the different mass estimates to the
manner in which a core is defined.  Automatic clump finding algorithms
like {\it clfind} appear to find large cores, if the core is embedded
in extended emission. Furthermore the previous studies which are both
at a single wavelength also identify ``clumps'', which appear
spurious, like some of the cores in the molecular cloud south of
NGC\,2023. A protostellar core should have a correspondence in some
dense gas tracers, like HCN, HCO$^+$, CS, etc. and stand out with a
single velocity. It can have a gradient due to collapse, rotation or
expansion.  All the cores do not necessarily have the same temperature
or dust properties. Cores with embedded protostars have rather
different dust properties ($\beta$-index) and temperature structure.
Resolution of the apparent disagreement between different sets of
observations will be possible only by using far-infrared observations
having similar spatial resolution, so that it is possible to determine
the dust emissivity and temperatures of these cores reliably. This
will soon be possible with missions like Herschel and SOFIA, both of
which have far-infrared imagers providing spatial resolution at a 100
$\mu$m of 8 -- 10\arcsec, very similar to what can be achieved on
large ground based single dish telescopes.  Only then can we reliably
assess how closely  the pre-stellar core mass function resembles the
interstellar mass function.

\section{Summary
\label{sec_summary}}

We have investigated the distribution of gas, dust and young stars in
the vicinity of NGC\,2023 and the L\,1630 molecular cloud using a
combination of the best-available datasets at multiple wavelengths. 

We reliably identify 5 sub-millimeter cores MM\,1--5 at 450 and
850~\micron\ using SCUBA observations. Of these cores MM\,1 and MM\,3
are bona fide Class I/0 sources. Single temperature greybody models
have been fitted to the SEDs of the sub-millimeter cores to derive
dust temperatures and emissivity indices ($\beta$). We find that
adjacent cores MM\,1 and MM\,2 show vastly different $\beta$ values of
0.7 and 2.1 respectively. Large differences in $\beta$ values are
obtained for the sources MM\,3 (0.8) and MM\,4 (1.5) as well. The
lower ($<1$) values of $\beta$ compared to those found in PMS objects
could be explained by the presence of partially optically thick disks
and/or grain growth in disks. In either case it is a clear signature
of the presence of protostellar disks. We find that the values of
$\beta$ correlate with the dust temperatures and hence the
evolutionary stage of the source. The derived significant variation of
$\beta$ values highlights a potential problem in the commonly made
assumption of uniform values of $\beta$ when calculating masses of
cores distributed over a region.

Using NIR (2MASS) and MIR ({\em Spitzer} IRAC and MIPS) photometry we
have generated color-color diagrams to estimate the ages of the
detected sources.  We identify a total of 73 PMS objects in the entire
22\arcmin\ $\times$ 26\arcmin\ region, out of which a possible 22--32
are associated with the very young cluster associated with NGC\,2023
of which HD\,37903 is the most massive member.  We categorize 5 YSOs
as Class 0/I objects and 10 sources as Class I/II objects. For the 10
sources which are detected only in the MIPS 24~\micron\ band the
protostellar nature needs to be confirmed.  Except for MM\,3 none of
the young MIR sources  has been detected in the sub-millimeter. This,
together with the fact that the available low (compared to the present
datasets) resolution FIR maps look significantly different from the
data presented here, emphasizes the need for follow-up FIR
observations with comparable resolution, in order to conclusively
determine the evolutionary stages of these objects.  The MIR datasets
also reveal several sources like HD37903, V615~Ori, Sellgren's star C
etc., which are bright in the optical. The UV datasets have  been used
to refine the spectral classification of HD37903 and the MIR datasets
further suggest that it as a Herbig Be star with significant infrared
excess. For V615~Ori, which most likely drives a weak extended CO
outflow, the infrared excess can be explained by a low-mass (0.06
\Msun{}) circumstellar disk and envelope, which is too faint to be
detected in the sub-millimeter.

We have used high resolution \nhtwod\ observations to probe the
protostellar disk associated with the Class 0 source MM\,1. Although we
detect enhanced \nhtwod\ emission, the velocities suggest that the
emission arises primarily from the surrounding ambient cloud which is
also rotating. Based on \CeiO\ observations we confirm CO depletion in
the MM\,1/MM\,2 region, consistent with the enhanced \nhtwod\ emission.
In contrast MM\,3, also identified to be a Class 0/I star and estimated
to have a dust temperature, somewhat higher than the temperature of
MM\,1/MM\,2 shows no signs of CO depletion.

Based on the derived distribution of YSOs in NGC~2023/L~1630 region we
conclude that the expansion of the  \HII\ region IC~434 has triggered
the formation of stars in the compressed gas inside the L~1630
molecular cloud, as well as in some of the dense ``elephant trunk"
structures. We have been able to derive reliable age estimates of most
of the YSOs in the mapped region based on NIR and MIR photometry.
However far-infrared observations at comparable angular resolutions
are absolutely essential to determine the evolutionary stages of the
extremely young objects which are clearly detected only at 24~\micron\
and also to confirm the protostellar nature of three of the
sub-millimeter clumps (MM\,2, MM\,4 and MM\,5) for which no other
signatures are so far available.

{\bf Acknowledgments}

We thank Tim Jenness for obtaining some of the SCUBA data discussed in
this paper. William Vacca helped us with spectral classification and
editing.  Brian Fleming provided us with a calibrated Spitzer IRS
spectrum of MM\,3.  This research used the facilities of the Canadian
Astronomy Data Centre operated by the National Research Council of
Canada with the support of the Canadian Space Agency. This paper has
made extensive use of  the SIMBAD database, operated at CDS,
Strasbourg, France. The National Radio Astronomy Observatory is a
facility of the National Science Foundation operated under cooperative
agreement by Associated Universities, Inc.

\addtocounter{table}{1}

\longtabL{2}{
{\tiny
\begin{landscape}
\begin{longtable}{lccrrrrrrrrll}
\caption{Coordinates and flux densities of MIR sources detected in the
vicinity of NGC~2023 
\label{tab_iracsrc}}\\
\hline\hline
Source & $\alpha_{2000}$ & $\delta_{2000}$ & $J$ & $H$ & $K_s$ & F$_{3.6~\micron}$ & F$_{4.5~\micron}$ & F$_{5.8~\micron}$ & F$_{8.0~\micron}$ & F$_{24~\micron}$ & Other Name & Class \\
L1630& & & & & & mJy & mJy & mJy & mJy & mJy & &\\
\hline
\endfirsthead
\caption{continued.}\\
\hline\hline
Source & $\alpha_{2000}$ & $\delta_{2000}$ & $J$ & $H$ & $K_s$ & F$_{3.6~\micron}$ & F$_{4.5~\micron}$ & F$_{5.8~\micron}$ & F$_{8.0~\micron}$ & F$_{24~\micron}$ & Other Name  & Class \\
L1630 && & & & & mJy & mJy & mJy & mJy & mJy &  & \\
\hline
\endhead
\hline
\endfoot
MIR-1 &  5:40:39.31 &  -2:26:45.8 & 11.58 $\pm$  0.03 & 10.70 $\pm$  0.02 & 10.31 $\pm$  0.02 &  28.71$\pm$0.10&     26.48 $\pm$ 0.07&     31.0 $\pm$ 0.6&     46.4 $\pm$ 0.4&     70.2 $\pm$ 0.2 &      &          II       \\
MIR-2 &  5:40:40.01 &  -2:12:46.7 & 12.20 $\pm$  0.03 & 11.45 $\pm$ 0.02 & 11.23 $\pm$  0.02 &  10.41$\pm$0.06&      6.14 $\pm$ 0.03& 5.3 $\pm$ 0.6&      2.3 $\pm$ 0.2& \ldots &      &          S       \\
MIR-3 &  5:40:41.39 &  -2:29:07.8 & 13.07 $\pm$  0.03 & 12.44 $\pm$ 0.03 & 12.15 $\pm$  0.03 &   5.45$\pm$0.02&      3.71 $\pm$ 0.02& 3.8 $\pm$ 0.5&      1.3 $\pm$ 0.2& \ldots &      &          S       \\
MIR-4 &  5:40:41.42 &  -2:19:11.2 & 10.29 $\pm$  0.03 &  9.61 $\pm$ 0.03 &  9.40 $\pm$  0.02 &  51.87$\pm$0.14&     32.27 $\pm$ 0.09& 23.2 $\pm$ 0.7&     13.7 $\pm$ 0.3& \ldots &      &          S       \\
MIR-5 &  5:40:44.56 &  -2:30:36.5 & 11.95 $\pm$  0.03 & 11.52 $\pm$ 0.02 & 11.36 $\pm$  0.02 &   8.19$\pm$0.03&      5.15 $\pm$ 0.02& 4.2 $\pm$ 0.4&      2.2 $\pm$ 0.2& \ldots &      &          S       \\
MIR-6 &  5:40:44.63 &  -2:22:43.9 & 11.82 $\pm$  0.03 & 11.14 $\pm$ 0.02 & 10.91 $\pm$  0.03 &  13.17$\pm$0.05&      8.73 $\pm$ 0.03& 7.8 $\pm$ 0.4&      3.4 $\pm$ 0.2& \ldots &      &          S       \\
MIR-7 &  5:40:44.75 &  -2:11:49.8 &  9.85 $\pm$  0.03 &  9.20 $\pm$  0.02 &  8.96 $\pm$  0.03 & 117.20$\pm$0.56&    102.20 $\pm$ 0.51&    104.5 $\pm$ 1.0&    112.4 $\pm$ 0.7&     74.2 $\pm$ 0.2 &      &          II       \\
MIR-8 &  5:40:44.75 &  -2:13:53.7 & 11.49 $\pm$  0.03 & 10.64 $\pm$ 0.03 & 10.39 $\pm$  0.02 &  21.82$\pm$0.07&     12.76 $\pm$ 0.04& 11.0 $\pm$ 0.4&      5.5 $\pm$ 0.2& \ldots &      &          S       \\
MIR-9 &  5:40:44.85 &  -2:12:15.6 &  9.76 $\pm$  0.03 &  9.00 $\pm$ 0.02 &  8.80 $\pm$  0.03 &  88.02$\pm$0.28&     56.26 $\pm$ 0.15& 42.8 $\pm$ 0.7&     24.1 $\pm$ 0.4&      3.2 $\pm$ 0.2 &      & S       \\
MIR-10 &  5:40:46.07 &  -2:28:03.3 & 12.08 $\pm$  0.03 & 11.42 $\pm$  0.02 & 11.24 $\pm$  0.03 &   9.46$\pm$0.03&      5.74 $\pm$ 0.02&      4.4 $\pm$ 0.4&      3.1 $\pm$ 0.2& \ldots &      &          II/III       \\
MIR-11 &  5:40:46.78 &  -2:10:49.9 & 11.93 $\pm$  0.03 & 11.02 $\pm$  0.03 & 10.51 $\pm$  0.03 &  30.03$\pm$0.11&     27.24 $\pm$ 0.07&     23.9 $\pm$ 0.6&     26.3 $\pm$ 0.3&     72.2 $\pm$ 0.2 &      &          II       \\
MIR-12 &  5:40:47.07 &  -2:24:48.6 & 11.51 $\pm$  0.03 & 11.17 $\pm$ 0.03 & 11.10 $\pm$  0.03 &  10.53$\pm$0.03&      6.97 $\pm$ 0.03& 5.4 $\pm$ 0.4&      2.1 $\pm$ 0.2& \ldots &      &          S       \\
MIR-13 &  5:40:47.53 &  -2:12:10.9 & 12.36 $\pm$  0.03 & 11.58 $\pm$ 0.02 & 11.36 $\pm$  0.03 &   8.40$\pm$0.04&      5.34 $\pm$ 0.02& 4.5 $\pm$ 0.5&      2.1 $\pm$ 0.2& \ldots &      &          S       \\
MIR-14 &  5:40:49.61 &  -2:28:56.0 &  9.70 $\pm$  0.03 &  8.98 $\pm$ 0.02 &  8.70 $\pm$  0.02 &  92.42$\pm$0.29&     58.86 $\pm$ 0.13& 42.7 $\pm$ 0.7&     24.0 $\pm$ 0.3&      3.2 $\pm$ 0.2 &      & S       \\
MIR-15 &  5:40:50.48 &  -2:12:53.9 &  9.88 $\pm$  0.03 &  8.99 $\pm$ 0.02 &  8.74 $\pm$  0.02 &  97.28$\pm$0.52&     65.38 $\pm$ 0.41& 46.5 $\pm$ 0.7&     22.7 $\pm$ 0.4& \ldots &      &          S       \\
MIR-16 &  5:40:50.94 &  -2:09:45.2 & 11.44 $\pm$  0.03 & 10.76 $\pm$ 0.02 & 10.53 $\pm$  0.02 &  20.54$\pm$0.07&     12.94 $\pm$ 0.04& 10.3 $\pm$ 0.5&      8.4 $\pm$ 0.3& \ldots &      &          II/III      \\
MIR-17 &  5:40:51.10 &  -2:20:59.8 & 11.30 $\pm$  0.03 & 11.02 $\pm$ 0.02 & 10.95 $\pm$  0.03 &  10.93$\pm$0.04&      7.49 $\pm$ 0.02& 6.2 $\pm$ 0.5&      2.9 $\pm$ 0.2& \ldots &      &          S       \\
MIR-18 &  5:40:51.73 &  -2:26:48.9 & 12.25 $\pm$  0.04 & 10.73 $\pm$  0.03 &  9.71 $\pm$  0.03 & 142.00$\pm$0.65&    166.40 $\pm$ 0.58&    217.4 $\pm$ 1.4&    322.3 $\pm$ 1.1&    754.5 $\pm$ 0.5 &  B33-1    &       II      \\
MIR-19 &  5:40:52.41 &  -2:27:12.6 & 15.09 $\pm$  0.05 & 13.61 $\pm$ 0.04 & 12.58 $\pm$  0.03 &   7.33$\pm$0.04&      9.12 $\pm$ 0.03& 12.7 $\pm$ 0.5&     22.2 $\pm$ 0.3&    105.0 $\pm$ 0.2 &  B33-28    &          I/II       \\
MIR-20 &  5:40:53.11 &  -2:15:48.4 & 11.02 $\pm$  0.03 & 10.51 $\pm$ 0.02 & 10.29 $\pm$  0.02 &  23.95$\pm$0.07&     15.91 $\pm$ 0.05& 11.9 $\pm$ 0.4&      6.1 $\pm$ 0.2& \ldots &      &          S       \\
MIR-21 &  5:40:53.75 &  -2:09:38.4 & 12.27 $\pm$  0.03 & 11.29 $\pm$ 0.02 & 10.86 $\pm$  0.02 &  15.85$\pm$0.05&     11.40 $\pm$ 0.03& 7.9 $\pm$ 0.4&      3.1 $\pm$ 0.2& \ldots &      &          S       \\
MIR-22 &  5:40:54.86 &  -2:24:06.6 & 12.04 $\pm$  0.03 & 11.27 $\pm$ 0.02 & 11.06 $\pm$  0.02 &  11.00$\pm$0.04&      7.28 $\pm$ 0.03& 6.6 $\pm$ 0.5&      3.1 $\pm$ 0.2& \ldots &      &          S       \\
MIR-23 &  5:40:56.09 &  -2:24:02.1 &  9.47 $\pm$  0.03 &  8.79 $\pm$ 0.05 &  8.57 $\pm$  0.02 & 101.30$\pm$0.52&     71.04 $\pm$ 0.35& 48.1 $\pm$ 0.6&     27.3 $\pm$ 0.3& 1.30$\pm$0.1 &      &          S       \\
MIR-24 &  5:40:56.68 &  -2:11:15.7 & 12.67 $\pm$  0.03 & 10.88 $\pm$  0.02 & 10.07 $\pm$  0.02 &  43.18$\pm$0.14&     29.22 $\pm$ 0.09&     22.7 $\pm$ 0.5&     15.3 $\pm$ 0.4& \ldots &      &          II/III       \\
MIR-25 &  5:40:56.73 &  -2:08:57.1 &  8.91 $\pm$  0.02 &  8.74 $\pm$ 0.04 &  8.64 $\pm$  0.02 &  97.09$\pm$0.51&     64.84 $\pm$ 0.36& 42.4 $\pm$ 0.7&     21.2 $\pm$ 0.4& \ldots &  HD\,290815    &          S       \\
MIR-26 &  5:40:56.89 &  -2:24:49.6 & 13.22 $\pm$  0.03 & 12.51 $\pm$  0.02 & 12.27 $\pm$  0.02 &   3.49$\pm$0.02&      2.27 $\pm$ 0.11&      1.9 $\pm$ 0.5&      1.0 $\pm$ 0.2& \ldots &      &          II/III       \\
MIR-27 &  5:40:58.14 &  -2:31:32.2 & 12.38 $\pm$  0.03 & 11.64 $\pm$ 0.03 & 11.46 $\pm$  0.02 &   8.41$\pm$0.03&      5.20 $\pm$ 0.02& 4.3 $\pm$ 0.4&      2.1 $\pm$ 0.2& \ldots &      &          S       \\
MIR-28 &  5:40:58.68 &  -2:25:26.7 & 12.25 $\pm$  0.03 & 10.77 $\pm$ 0.02 & 10.06 $\pm$  0.02 &  39.42$\pm$0.14&     30.27 $\pm$ 0.09& 21.2 $\pm$ 0.6&     13.5 $\pm$ 0.3& \ldots & B33-10     &          II/III       \\
MIR-29 &  5:41:01.57 &  -2:21:13.9 & 10.47 $\pm$  0.03 &  8.81 $\pm$ 0.03 &  8.05 $\pm$  0.03 & 218.80$\pm$1.00&    143.40 $\pm$ 0.64& 113.7 $\pm$ 0.9&     63.6 $\pm$ 0.6& \ldots &      &          S       \\
MIR-30 &  5:41:01.92 &  -2:07:03.4 & 13.96 $\pm$  0.03 & 12.21 $\pm$  0.02 & 11.34 $\pm$  0.02 &  13.30$\pm$0.06&     10.18 $\pm$ 0.04&      7.4 $\pm$ 0.5&      3.5 $\pm$ 0.2& \ldots &      &          II/III       \\
MIR-31 &  5:41:02.67 &  -2:18:17.8 &  6.95 $\pm$  0.03 &  6.87 $\pm$ 0.05 &  6.76 $\pm$  0.02 & 517.60$\pm$1.86&    337.60 $\pm$ 1.18& 228.3 $\pm$ 1.2&    131.0 $\pm$ 0.7&     55.4 $\pm$ 0.2 &   HD\,37805   & S       \\
MIR-32 &  5:41:03.57 &  -2:25:12.5 & 11.81 $\pm$  0.03 & 10.86 $\pm$ 0.03 & 10.38 $\pm$  0.02 &  23.72$\pm$0.08&     16.13 $\pm$ 0.05& 11.8 $\pm$ 0.5&      7.7 $\pm$ 0.3& \ldots &      &          S       \\
MIR-33 &  5:41:03.78 &  -2:06:26.3 & 12.85 $\pm$  0.03 & 12.25 $\pm$  0.02 & 11.94 $\pm$  0.02 &   7.30$\pm$0.03&      5.90 $\pm$ 0.03&      5.7 $\pm$ 0.4&      4.8 $\pm$ 0.3&     15.0 $\pm$ 0.2 &      &          II       \\
MIR-34 &  5:41:05.44 &  -2:31:05.9 & 12.74 $\pm$  0.03 & 11.75 $\pm$  0.03 & 11.40 $\pm$  0.02 &   9.72$\pm$0.03&      5.90 $\pm$ 0.02&      6.0 $\pm$ 0.4&      6.9 $\pm$ 0.3& \ldots &      &          II/III       \\
MIR-35 &  5:41:05.44 &  -2:24:59.0 &   \ldots & \ldots &  \ldots &  \ldots & \ldots & 1.3$\pm$ 0.4   &   3.2 $\pm$ 0.3& \ldots &        &   II?               \\
MIR-36 &  5:41:09.41 &  -2:14:34.0 & 11.79 $\pm$  0.03 & 10.96 $\pm$ 0.02 & 10.60 $\pm$  0.02 &  19.74$\pm$0.06&     13.60 $\pm$ 0.05& 11.1 $\pm$ 0.5&      5.5 $\pm$ 0.3& \ldots &      &          S       \\
MIR-37 &  5:41:10.31 &  -2:29:04.0 & 13.83 $\pm$ 0.03 & 13.27$\pm$0.03 & 12.91$\pm$0.03 & 2.39$\pm$0.02 & 1.83$\pm$0.01 & 1.27$\pm$0.03 & 1.22$\pm$0.2 & \ldots & B33-21 & II/III\\
MIR-38 &  5:41:12.23 &  -2:27:34.5 & 13.70 $\pm$  0.03 & 12.12 $\pm$ 0.03 & 11.39 $\pm$  0.03 &  12.08$\pm$0.05&      7.77 $\pm$ 0.03& 6.8 $\pm$ 0.4&      4.6 $\pm$ 0.2& \ldots &   B33-22   &          II/III       \\
MIR-39 &  5:41:12.40 &  -2:19:21.1 & \ldots            & \ldots            & \ldots            &  0.35$\pm$0.01&      0.35 $\pm$ 0.01&      1.4 $\pm$ 0.4&      2.6 $\pm$ 0.2& \ldots &      &          I/II       \\
MIR-40 &  5:41:13.06 &  -2:29:12.2 & 11.78 $\pm$  0.03 & 11.37 $\pm$ 0.02 & 11.21 $\pm$  0.02 &   9.22$\pm$0.03&      6.42 $\pm$ 0.02& 4.7 $\pm$ 0.4&      2.2 $\pm$ 0.2& \ldots &  B33-23    &          S       \\
MIR-41 &  5:41:14.06 &  -2:25:05.1 & 12.76 $\pm$  0.03 & 12.11 $\pm$ 0.03 & 11.91 $\pm$  0.02 &   5.78$\pm$0.24&      3.57 $\pm$ 0.02& 3.4 $\pm$ 0.1  &      4.1 $\pm$ 0.2& \ldots &  B33-25    &          II      \\
MIR-42 &  5:41:14.45 &  -2:29:35.4 & 10.81 $\pm$  0.03 & 10.51 $\pm$ 0.03 & 10.43 $\pm$  0.02 &  17.24$\pm$0.06&     12.20 $\pm$ 0.04& 8.7 $\pm$ 0.5&      3.9 $\pm$ 0.2& \ldots &      &          S       \\
MIR-43 &  5:41:17.18 &  -2:18:07.3 & \ldots & \ldots & \ldots & 0.35$\pm$0.00&      0.67 $\pm$ 0.01&      2.2 $\pm$ 0.4&      3.0 $\pm$ 0.2&     14.9 $\pm$ 0.2 &      &          I/II       \\
MIR-44 &  5:41:17.33 &  -2:31:38.1 &  9.64 $\pm$  0.03 &  9.32 $\pm$ 0.03 &  9.18 $\pm$  0.02 &      60.57$\pm$0.19&     38.14 $\pm$ 0.10& 25.6 $\pm$ 0.6&     13.8 $\pm$ 0.3& \ldots &      &           S       \\
MIR-45 &  5:41:20.89 &  -2:17:52.8 & 15.43 $\pm$  0.05 & 13.35 $\pm$ 0.02 & 12.16 $\pm$  0.02 &      10.61$\pm$0.04&     12.62 $\pm$ 0.04& 13.0 $\pm$ 0.5&     13.8 $\pm$ 0.3& 2.0 $\pm$ 1.0&      & II/III       \\
MIR-46 &  5:41:21.70 &  -2:11:08.2 & \ldots & \ldots & \ldots & 3.53$\pm$0.02&      4.84 $\pm$ 0.21&      7.8 $\pm$ 0.4&      9.1 $\pm$ 0.3&     23.4 $\pm$ 0.2 & VLA\,3     &           II       \\
MIR-47 &  5:41:22.14 &  -2:16:44.0 & 15.62 $\pm$  0.06 & 13.58 $\pm$ 0.03 & 12.48 $\pm$  0.03 &       8.50$\pm$0.33&      8.59 $\pm$ 0.03& 9.3 $\pm$ 0.5&      8.4 $\pm$ 0.3& 2.0 $\pm$ 1.0 &      & II/III       \\
MIR-48 &  5:41:22.48 &  -2:17:18.6 & 14.68 $\pm$  0.04 & 12.95 $\pm$  0.03 & 12.05 $\pm$  0.02 &       8.05$\pm$0.32&      6.89 $\pm$ 0.03&      5.3 $\pm$ 0.4&      2.8 $\pm$ 0.2& \ldots &      &           II/III       \\
MIR-49 &  5:41:22.49 &  -2:09:14.7 & 16.93 $\pm$ \ldots & 15.74 $\pm$ 0.13 & 14.60 $\pm$  0.11 &       0.74$\pm$0.01&      0.69 $\pm$ 0.01& 1.2 $\pm$ 0.5&      3.8 $\pm$ 0.2&     13.2 $\pm$ 0.1 &      & I/II       \\
MIR-50 &  5:41:23.29 &  -2:17:35.7 & 17.12 $\pm$ \ldots & 16.53 $\pm$ \ldots & 14.25 $\pm$  0.07 &       4.57$\pm$0.02&      8.59 $\pm$ 0.03&     14.8 $\pm$ 0.5&     19.8 $\pm$ 0.4&    122.6 $\pm$ 0.2 &      &           I         \\
MIR-51 &  5:41:24.21 &  -2:16:06.5 & \ldots & \ldots & \ldots & 2.61$\pm$0.01&      5.63 $\pm$ 0.02&      5.3 $\pm$ 0.4&      5.0 $\pm$ 0.3&    369.8 $\pm$ 0.3 &      &           I/II       \\
MIR-52 &  5:41:24.49 &  -2:22:36.1 &  9.10 $\pm$  0.02 &  8.07 $\pm$  0.04 &  7.39 $\pm$  0.02 &     922.90$\pm$2.29&    900.90 $\pm$ 2.44&    980.1 $\pm$ 2.8&    897.9 $\pm$ 3.0&   1067.0 $\pm$ 0.5 &   V615~Ori   &           II       \\
MIR-53 &  5:41:26.20 &  -2:10:07.5 & 10.46 $\pm$  0.03 & 10.22 $\pm$ 0.02 & 10.13 $\pm$  0.02 &      22.49$\pm$0.08&      6.20 $\pm$ 0.03& 10.7 $\pm$ 0.5&      6.3 $\pm$ 0.2& \ldots &      &           S       \\
MIR-54 &  5:41:26.23 &  -2:16:22.3 & \ldots & \ldots & \ldots &  \ldots& \ldots&      1.9 $\pm$ 0.4&      4.8 $\pm$ 0.3& \ldots &      & II?      \\
MIR-55 &  5:41:27.98 &  -2:19:40.2 & 11.34 $\pm$  0.02 & 10.94 $\pm$ 0.02 & 10.89 $\pm$  0.02 &      11.45$\pm$0.05&      7.88 $\pm$ 0.03& 6.3 $\pm$ 0.4&      3.3 $\pm$ 0.2& \ldots &      &           S       \\
MIR-56 &  5:41:28.95 &  -2:23:19.2 & \ldots & \ldots & \ldots &       0.27$\pm$0.00&      1.14 $\pm$ 0.01&      3.5 $\pm$ 0.4&      4.6 $\pm$ 0.2&     91.6 $\pm$ 0.2 &      &           I       \\
MIR-57 &  5:41:30.24 &  -2:07:31.8 & 13.81 $\pm$  0.03 & 13.18 $\pm$  0.03 & 12.89 $\pm$  0.03 &       2.42$\pm$0.01&      1.94 $\pm$ 0.01&      2.7 $\pm$ 0.4&      3.5 $\pm$ 0.2&     25.1 $\pm$ 0.1 &      &           II       \\
MIR-58 &  5:41:32.95 &  -2:28:02.8 & 15.09 $\pm$  0.03 & 12.95 $\pm$ 0.03 & 11.92 $\pm$  0.02 &       9.80$\pm$0.04&      7.44 $\pm$ 0.03& 6.5 $\pm$ 0.4&      3.0 $\pm$ 0.3& \ldots &      &           II/III       \\
MIR-59 &  5:41:33.90 &  -2:19:00.3 & 18.61 $\pm$ \ldots & 15.43 $\pm$ 0.08 & 13.09 $\pm$  0.03 &       5.73$\pm$0.02&      5.52 $\pm$ 0.02& 5.4 $\pm$ 0.4&      3.3 $\pm$ 0.2& \ldots &      &           II/III       \\
MIR-60 &  5:41:34.72 &  -2:17:24.0 & 12.28 $\pm$  0.03 & 11.42 $\pm$  0.02 & 10.97 $\pm$  0.02 &      19.50$\pm$0.08&     17.89 $\pm$ 0.06&     17.2 $\pm$ 0.5&     20.4 $\pm$ 0.4& \ldots &  E/1    &           II       \\
MIR-61 &  5:41:36.01 &  -2:08:21.1 & 14.76 $\pm$  0.04 & 13.38 $\pm$ 0.03 & 12.73 $\pm$  0.02 &       3.10$\pm$0.02&      2.32 $\pm$ 0.01& 1.8 $\pm$ 0.5&      1.2 $\pm$ 0.2& \ldots &      &           S       \\
MIR-62 &  5:41:36.39 &  -2:16:46.1 & 11.73 $\pm$  0.03 & 10.05 $\pm$ 0.02 &  8.83 $\pm$  0.02 &     505.40$\pm$1.87&    622.50 $\pm$ 1.66& 667.3 $\pm$ 2.9&    815.8 $\pm$ 3.2&   1063.0 $\pm$ 1.0 &  C/2/WB54    & I/II       \\
MIR-63 &  5:41:37.18 &  -2:17:17.2 & 16.94 $\pm$ \ldots & 13.86 $\pm$  0.07 & 10.69 $\pm$ \ldots &     212.50$\pm$0.77&    520.50 $\pm$ 1.60&   1031.0 $\pm$ 3.2&   1791.0 $\pm$ 5.4&  25604 $\pm$ 640 & D/3/MM\,3     &           I       \\
MIR-64 &  5:41:37.46 &  -2:30:21.1 & \ldots & \ldots & \ldots &
0.22 $\pm$ 0.04&  0.22 $\pm$ 0.04  & 0.29$\pm$0.02 & 2.4$\pm$0.2 & \ldots & &         II?       \\
MIR-65 &  5:41:37.90 &  -2:25:27.2 & 14.15 $\pm$  0.03 & 11.46 $\pm$ 0.02 & 10.18 $\pm$  0.02 &      45.88$\pm$0.14&     33.72 $\pm$ 0.09& 28.5 $\pm$ 0.6&     15.0 $\pm$ 0.3& \ldots &      &           S       \\
MIR-66 &  5:41:38.40 &  -2:15:32.7 &  7.37 $\pm$  0.02 &  7.42 $\pm$ 0.05 &  7.28 $\pm$  0.02 &     304.40$\pm$1.29&    241.20 $\pm$ 0.78& 182.5 $\pm$ 1.5&    144.1 $\pm$ 1.4& 4500 $\pm$ 2000 &   HD\,37903   & I/II       \\
MIR-67 &  5:41:40.25 &  -2:15:59.7 & 11.31 $\pm$  0.03 & 10.36 $\pm$  0.02 &  9.89 $\pm$  0.02 &      23.35$\pm$0.18&     41.00 $\pm$ 0.15&     53.1 $\pm$ 1.0&    100.6 $\pm$ 1.5& \ldots & B/5/WB55     &     I/II       \\
MIR-68 &  5:41:42.23 &  -2:16:24.5 & 15.96 $\pm$  0.11 & 12.84 $\pm$ 0.07 & 10.91 $\pm$  0.04 &      48.04$\pm$0.19&     56.38 $\pm$ 0.17& 75.3 $\pm$ 0.7&     93.1 $\pm$ 0.9&    602.9 $\pm$ 0.4 & I/7     & I/II  \\
MIR-69 &  5:41:42.25 &  -2:17:35.9 & 12.31 $\pm$  0.02 & 11.48 $\pm$  0.02 & 10.99 $\pm$  0.02 &      23.17$\pm$0.09&     23.70 $\pm$ 0.07&     25.9 $\pm$ 0.6&     25.5 $\pm$ 0.4&     54.4 $\pm$ 0.2 &  G/9/WB56    &           II       \\
MIR-70 &  5:41:42.39 &  -2:17:21.0 & 14.57 $\pm$  0.03 & 13.20 $\pm$  0.02 & 12.47 $\pm$  0.02 &       6.06$\pm$0.03&      6.20 $\pm$ 0.03&      7.7 $\pm$ 0.5&      7.8 $\pm$ 0.4&    101.7 $\pm$ 0.2 &    8  &    II     \\
MIR-71 &  5:41:43.10 &  -2:10:26.1 & 11.25 $\pm$  0.02 & 10.63 $\pm$ 0.02 & 10.49 $\pm$  0.02 &      19.05$\pm$0.07&     11.70 $\pm$ 0.03& 8.9 $\pm$ 0.5&      4.3 $\pm$ 0.2& \ldots &      &           S       \\
MIR-72 &  5:41:44.23 &  -2:16:16.3 & 12.86 $\pm$  0.02 & 12.04 $\pm$  0.02 & 11.50 $\pm$  0.02 &      12.99$\pm$0.08&     11.65 $\pm$ 0.05&     13.2 $\pm$ 0.5&     14.5 $\pm$ 0.5& \ldots &  12/W217/WB58    &            II       \\
MIR-73 &  5:41:44.62 &  -2:16:06.8 & 15.19 $\pm$  0.06 & 12.43 $\pm$  0.08 & 10.38 $\pm$  0.03 &     311.00$\pm$1.00&    537.50 $\pm$ 1.61&    691.5 $\pm$ 2.3&    729.8 $\pm$ 2.3&  11272 $\pm$ 168 &  14/W219   &           I       \\
MIR-74 &  5:41:44.69 &  -2:16:56.6 & 13.21 $\pm$  0.03 & 12.12 $\pm$  0.03 & 11.54 $\pm$  0.03 &       9.52$\pm$0.05&      7.96 $\pm$ 0.04&      7.8 $\pm$ 0.5&      6.3 $\pm$ 0.4& \ldots & 11/W211     &           II       \\
MIR-75 &  5:41:44.79 &  -2:15:55.1 & 13.75 $\pm$  0.08 & 11.44 $\pm$  0.05 &  9.99 $\pm$  0.03 &     137.90$\pm$0.53&    231.20 $\pm$ 0.89&    279.1 $\pm$ 1.4&    310.3 $\pm$ 1.5&   1439.0 $\pm$ 0.6 &  F/15/W220    &           I       \\
MIR-76 &  5:41:45.50 &  -2:24:16.0 & 13.19 $\pm$  0.03 & 11.52 $\pm$ 0.02 & 10.80 $\pm$  0.02 &      18.74$\pm$0.06&     13.95 $\pm$ 0.04& 10.9 $\pm$ 0.4&      5.7 $\pm$ 0.2& \ldots &      &           S       \\
MIR-77 &  5:41:45.69 &  -2:24:36.3 & 12.61 $\pm$  0.02 & 11.46 $\pm$ 0.02 & 10.87 $\pm$  0.02 &      16.69$\pm$0.05&     11.87 $\pm$ 0.04& 9.2 $\pm$ 0.4&      4.6 $\pm$ 0.2& \ldots &      &           S       \\
MIR-78 &  5:41:46.08 &  -2:31:42.8 & 12.77 $\pm$  0.02 & 11.64 $\pm$ 0.02 & 11.20 $\pm$  0.02 &      10.83$\pm$0.04&      6.87 $\pm$ 0.03& 6.2 $\pm$ 0.5&      2.8 $\pm$ 0.2& \ldots &      &           S       \\
MIR-79 &  5:41:46.68 &  -2:32:05.1 & 13.53 $\pm$  0.02 & 12.45 $\pm$  0.02 & 11.91 $\pm$  0.02 &       7.61$\pm$0.03&      6.55 $\pm$ 0.03&      6.2 $\pm$ 0.5&      5.4 $\pm$ 0.3&     11.4 $\pm$ 0.2 &      &            II       \\
MIR-80 &  5:41:47.04 &  -2:16:37.8 &  9.04 $\pm$  0.02 &  8.14 $\pm$  0.04 &  7.55 $\pm$  0.03 &     691.70$\pm$2.12&    721.70 $\pm$ 2.03&    640.6 $\pm$ 2.4&    657.5 $\pm$ 2.4&    458.7 $\pm$ 0.3 & H/16    &           II       \\
MIR-81 &  5:41:48.32 &  -2:10:40.4 & 10.89 $\pm$  0.03 & 10.62 $\pm$ 0.02 & 10.58 $\pm$  0.02 &      15.52$\pm$0.06&     10.53 $\pm$ 0.03& 7.8 $\pm$ 0.5&      4.2 $\pm$ 0.2& \ldots &      &           S       \\
MIR-82 &  5:41:50.35 &  -2:31:00.8 & 13.09 $\pm$  0.02 & 12.15 $\pm$ 0.03 & 11.78 $\pm$  0.02 &      5.81$\pm$0.03&     2.69 $\pm$ 0.03& 1.5 $\pm$ 0.2&      4.2 $\pm$ 0.2& \ldots &      &           S       \\
MIR-83 &  5:41:51.40 &  -2:29:51.1 &  8.90 $\pm$  0.02 &  8.60 $\pm$ 0.02 &  8.45 $\pm$  0.02 &     131.60$\pm$0.61&     85.62 $\pm$ 0.46& 58.8 $\pm$ 0.8&     31.0 $\pm$ 0.3&      2.4 $\pm$ 0.2 &      & S       \\
MIR-84 &  5:41:51.60 &  -2:07:41.3 & 14.01 $\pm$ \ldots & 13.24 $\pm$ 0.04 & 12.52 $\pm$  0.04 &       4.12$\pm$0.02&      2.91 $\pm$ 0.02& 2.9 $\pm$ 0.5&      1.3 $\pm$ 0.2& \ldots &      &           S       \\
MIR-85 &  5:41:52.32 &  -2:18:32.5 & 12.90 $\pm$  0.02 & 11.05 $\pm$ 0.02 & 10.16 $\pm$  0.02 &      36.53$\pm$0.11&     24.32 $\pm$ 0.07& 19.8 $\pm$ 0.5&     10.7 $\pm$ 0.3& \ldots &      &           S       \\
MIR-86 &  5:41:55.57 &  -2:23:40.3 & 12.85 $\pm$  0.02 & 11.66 $\pm$  0.02 & 11.05 $\pm$  0.02 &      25.95$\pm$0.10&     27.56 $\pm$ 0.06&     29.9 $\pm$ 0.6&     35.0 $\pm$ 0.3&    120.8 $\pm$ 0.3 &      &           II       \\
MIR-87 &  5:41:56.32 &  -2:12:45.6 & 12.31 $\pm$  0.02 & 11.15 $\pm$ 0.02 & 10.61 $\pm$  0.02 &      18.88$\pm$0.06&     12.86 $\pm$ 0.04& 10.3 $\pm$ 0.5&      5.3 $\pm$ 0.3& \ldots &      &           S       \\
MIR-88 &  5:41:57.94 &  -2:11:24.2 & 12.09 $\pm$  0.03 & 10.61 $\pm$ 0.02 & 10.03 $\pm$  0.02 &      31.79$\pm$0.07&     24.88 $\pm$ 0.07& 19.3 $\pm$ 0.4&     11.1 $\pm$ 0.2& \ldots &      &           S       \\
MIR-89 &  5:41:59.30 &  -2:17:02.1 & 13.22 $\pm$  0.02 & 11.67 $\pm$ 0.02 & 11.02 $\pm$  0.02 &      15.61$\pm$0.06&      9.45 $\pm$ 0.03& 9.4 $\pm$ 0.5&      4.9 $\pm$ 0.2& \ldots &      &           S       \\
MIR-90 &  5:41:59.90 &  -2:21:08.7 &  9.99 $\pm$  0.02 &  9.00 $\pm$ 0.03 &  8.56 $\pm$  0.02 &     124.20$\pm$0.49&     82.14 $\pm$ 0.46& 63.1 $\pm$ 0.7&     35.0 $\pm$ 0.4&      1.7 $\pm$ 0.1 &      & S       \\
MIR-91 &  5:42:00.01 &  -2:31:48.8 & 12.22 $\pm$  0.02 & 11.23 $\pm$ 0.02 & 10.88 $\pm$  0.02 &      14.55$\pm$0.06&      9.45 $\pm$ 0.03& 6.8 $\pm$ 0.5&      3.5 $\pm$ 0.2& \ldots &  Kiso-A    &           S       \\
MIR-92 &  5:42:02.31 &  -2:23:15.6 & 10.67 $\pm$  0.02 & 10.37 $\pm$ 0.02 & 10.28 $\pm$  0.02 &      21.45$\pm$0.06&     13.84 $\pm$ 0.05& 10.5 $\pm$ 0.4&      5.9 $\pm$ 0.2& \ldots &      &           S       \\
MIR-93 &  5:42:02.34 &  -2:23:28.4 & 12.70 $\pm$  0.02 & 11.88 $\pm$ 0.02 & 11.48 $\pm$  0.02 &       8.62$\pm$0.03&      5.80 $\pm$ 0.03& 5.1 $\pm$ 0.4&      3.0 $\pm$ 0.2& \ldots &      &           S       \\
MIR-94 &  5:42:02.64 &  -2:07:45.8 & \ldots & \ldots & \ldots & 0.51$\pm$0.00&      2.88 $\pm$ 0.01&      7.5 $\pm$ 0.4&      7.3 $\pm$ 0.2&     64.3 $\pm$ 0.2 &      &           I/II       \\
MIR-95 &  5:42:03.77 &  -2:24:50.4 & 13.15 $\pm$  0.02 & 11.81 $\pm$ 0.02 & 11.28 $\pm$  0.02 &      11.01$\pm$0.04&      7.00 $\pm$ 0.03& 5.4 $\pm$ 0.4&      3.4 $\pm$ 0.2& \ldots &      &           S       \\
\end{longtable}
$^{a}$ In column 12, integers correspond
to NIR sources detected by \citet{depoy1990}, alphabet B--F are for
sources identified by \citet{sellgren1983}, WB stands for sources
detected by \citet{weaver2004}, W are sources detected by
\citet{witt1984} and B33 names are from \citet{bowler2009}
\end{landscape}
}}

\end{document}